\documentclass[aps,prx,twocolumn,superscriptaddress,a4paper,showpacs]{revtex4}

\usepackage{graphicx}
\usepackage{amsmath}
\usepackage{float}
\usepackage{color}

\pdfoutput=1 

\usepackage{fancyhdr}
\usepackage[colorlinks=true,citecolor=blue,linkcolor=magenta]{hyperref}
\hypersetup{
pdfauthor = {Daniel Riedel},
colorlinks = true, linkcolor = blue, urlcolor=blue, bookmarksnumbered =  true}

\begin{document}
\title{Deterministic enhancement of coherent photon generation from a nitrogen-vacancy center in ultrapure diamond}

\author{Daniel Riedel}
\email{daniel.riedel@unibas.ch}
\affiliation{Department of Physics, University of Basel, Klingelbergstrasse 82, Basel CH-4056, Switzerland}
\author{Immo S\"ollner}
\affiliation{Department of Physics, University of Basel, Klingelbergstrasse 82, Basel CH-4056, Switzerland}
\author{Brendan J. Shields}
\affiliation{Department of Physics, University of Basel, Klingelbergstrasse 82, Basel CH-4056, Switzerland}
\author{Sebastian Starosielec}
\affiliation{Department of Physics, University of Basel, Klingelbergstrasse 82, Basel CH-4056, Switzerland}
\author{Patrick Appel}
\affiliation{Department of Physics, University of Basel, Klingelbergstrasse 82, Basel CH-4056, Switzerland}
\author{Elke Neu}
\altaffiliation{Current address: Department of Physics, Saarland University, Campus E2 6, Saarbr\"ucken DE-66123, Germany}
\affiliation{Department of Physics, University of Basel, Klingelbergstrasse 82, Basel CH-4056, Switzerland}
\author{Patrick Maletinsky}
\affiliation{Department of Physics, University of Basel, Klingelbergstrasse 82, Basel CH-4056, Switzerland}
\author{Richard J. Warburton}
\affiliation{Department of Physics, University of Basel, Klingelbergstrasse 82, Basel CH-4056, Switzerland}


\date{\today}

\begin{abstract}
The nitrogen-vacancy (NV) center in diamond has an optically addressable, highly coherent spin. However, an NV center even in high quality single-crystalline material is a very poor source of single photons: extraction out of the high-index diamond is inefficient, the emission of coherent photons represents just a few per cent of the total emission, and the decay time is large. In principle, all three problems can be addressed with a resonant microcavity. In practice, it has proved difficult to implement this concept: photonic engineering hinges on nano-fabrication yet it is notoriously difficult to process diamond without degrading the NV centers. We present here a microcavity scheme which uses minimally processed diamond, thereby preserving the high quality of the starting material, and a tunable microcavity platform. We demonstrate a clear change in the lifetime for multiple individual NV centers on tuning both the cavity frequency and anti-node position, a Purcell effect. The overall Purcell factor $F_{\rm P}=2.0$ translates to a Purcell factor for the zero phonon line (ZPL) of $F_{\rm P}^{\rm ZPL}\sim30$ and an increase in the ZPL emission probability from $\sim 3$\,\% to $\sim 46$\,\%. By making a step-change in the NV's optical properties in a deterministic way, these results pave the way for much enhanced spin-photon and spin-spin entanglement rates.
\end{abstract}

\pacs{42.55.Sa, 42.50.-p, 42.60.Da, 61.72.jn}

\maketitle

The nitrogen-vacancy (NV) center in diamond constitutes a workhorse in quantum technology on account of its optically addressable, coherent electron spin\,\cite{Jelezko2004}.
The NV stands out for its long spin coherence times\,\cite{Balasubramanian2009}, robust single photon emission\,\cite{Kurtsiefer2000} and the possibility of mapping its spin state to nearby nuclear spins\,\cite{Dutt2007}. Advances in spin-photon entanglement\,\cite{Togan2010} and two-photon quantum interference protocols\,\cite{Bernien2012,Sipahigil2012} pave the way for the implementation of quantum teleportation\,\cite{Pfaff2014} and long-distance spin-spin entanglement\,\cite{Hensen2015}. However, the success rate of these protocols and the scaling up to extended networks are both limited by the very small generation rate of indistinguishable photons from individual NV centers\,\cite{Gao2015}.\\
\indent There are at least four factors which limit the generation rate of indistinguishable photons. First, the lifetime of NV centers is relatively long, $\sim 12$\,ns. Secondly, only a small fraction, \mbox{$\sim3-4$\,\%}, of the NV emission goes into the zero phonon line (ZPL)\,\cite{Barclay2011,Kaupp2013}. Only ZPL emission is useful for photon-based entanglement-swapping protocols as the phonon involved in non-ZPL emission dephases very rapidly. Thirdly, the photon extraction efficiency out of the diamond is hindered by the large refractive index of diamond itself. Finally, there are random spectral fluctuations in the exact frequency of the NV emission caused by charge noise in the diamond host\,\cite{Bernien2012}.

Coupling the NV center to a high quality factor, low mode volume optical microcavity offers a potential remedy to the first three factors thereby dramatically improving the rate of coherent photon generation. These improvements depend on the weak coupling regime of cavity quantum electro-dynamics in which the emitter couples irreversibly to a single microcavity mode. The microcavity increases the total emission rate, the Purcell effect, and on resonance with the ZPL, the fraction of emission into the ZPL is likewise increased. The same coupling also enhances the ZPL extraction efficiency: photons leaking out of the cavity are channeled into a single propagating mode. A notable feature is that, as compared to the strong coupling regime of cavity quantum electrodynamics, modest cavity performance is all that is required\,\cite{Khitrova2006}.
 \begin{figure}[t]
\includegraphics[width=0.48\textwidth]{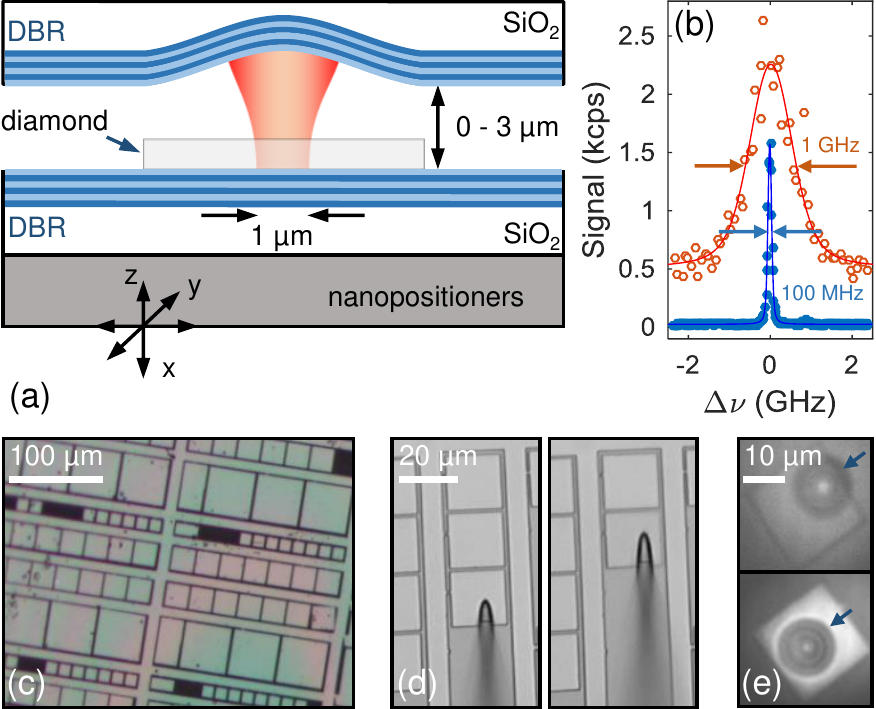}
\caption{(a) Schematic of the tunable microcavity containing a thin diamond membrane. Both the antinode location and resonance frequency of the microcavity mode can be tuned {\em in situ}. (b) Photoluminescence excitation (PLE) scans of near-surface (68\,nm) NV centers in unprocessed diamond (filled blue circles, $P_{\rm red}=3\,$nW) and in microstructured (${t_{\rm d} \leq 1\,\mu}$m) diamond (open red circles, $P_{\rm red}=100\,$nW) yielding zero-phonon linewidths of $\sim100$\,MHz and $\sim1$\,GHz, respectively. (c) Micro-structured diamond. (d) Detaching a $20 \times 20$\,$\mu$m$^{2}$ membrane using a micromanipulator. (e) Images (recorded with a wavelength out with the DBR stopband) of the diamond--microcavity demonstrating the {\em in situ} control of the lateral position. The arrows indicate the position of the concave top mirror.}
\label{fig:schematic}
\end{figure}
Implementing these concepts for emitters in diamond has proved difficult so far. Nevertheless, coupling of NVs to photonic crystal cavities has shown to convey significant improvements on account of the particularly small mode volume \mbox{$V$\,\cite{Sipahigil2016,Riedrich2015,Faraon2012,Hausmann2013,Li2014}}. However, fabricating such structures exhibiting high quality factors ($Q$-factors) in diamond is challenging: diamond is a very hard and chemically inert material. Furthermore, achieving a spatial and spectral resonance with a single emitter is also difficult such that device yield is poor. Also, efficient out-coupling is hard to engineer. In addition, the invasive processing causes a worsening of the spectral fluctuations, particularly for NV centers, the fourth problem mentioned above. 

In comparison, a miniaturized Fabry-P\'{e}rot microcavity has the advantage of {\em in situ} spatial and spectral tuning, along with high $Q$-factors and good mode-matching to a propagating Gaussian beam, at the expense of an increased mode volume\,\cite{Barbour2011,Greuter2014,Greuter2015}. The feasibility of this approach has been demonstrated by enhancing the emission rate of emitters in nanocrystals\,\cite{Albrecht2013,Johnson2015,Kaupp2016,Benedikter2017}. However, as in photonic crystal cavities, NVs in nanocrystals typically suffer from significant line broadening due to their close proximity to fluctuating charges at the surface. These spectral fluctuations are so severe that schemes involving photon-based entanglement swapping have only been successfully implemented using high-purity single-crystalline diamond material\,\cite{Greentree2006, Schroder2016}.

We present here deterministic enhancement of the ZPL emission rate from single NV centers with narrow ZPL linewidths ($\sim 1$\,GHz) by resonant coupling to a high-$Q$ microcavity mode. We demonstrate an increase of the probability of ZPL emission to $\sim 46$\,\%. Two principles have guided our work. First, at this stage of diamond-based quantum technology, {\em in situ} tuning of both microcavity frequency and anti-node position is extremely valuable. We have therefore opted for a miniaturized Fabry-P\'{e}rot microcavity [Fig.\,\ref{fig:schematic}(a)]. Secondly, we use ultrapure diamond material with minimal processing in order to ensure good NV optical properties. Specifically, thin diamond membranes are created out of high purity, single-crystalline chemical vapor deposition (CVD) diamond.
As starting material, we employ commercially available CVD diamond (Element 6, (100)-orientation) and introduce NV centers at a target depth of 68\,nm by nitrogen implantation ($^{14}$N, 55\,keV, ${2\cdot10^9}$\,ions/cm$^2$, straggle 16\,nm). Using multi-step high-temperature annealing, NV centers with close to lifetime-limited emission linewidths can be created\,\cite{Chu2014b}. Here, photoluminescence excitation (PLE) scans of NVs in the starting material yield ZPL linewidths of $\lesssim 100$\,MHz at 4\,K [Fig.\,\ref{fig:schematic}(b)]. Membranes of thickness $t_{\rm d} \lesssim 1$\,$\mu$m (with typical lateral dimensions $20 \times 20$\,$\mu$m$^{2}$) and surface roughness of $\lesssim 0.3$\,nm are fabricated from this starting material by plasma etching and microstructuring\,\cite{Appel2016, Maletinsky2012,Riedel2014} [Fig.\,\ref{fig:schematic}(c)]. Using a micro-manipulator, we break out membranes [Fig.\,\ref{fig:schematic}(d)] and transfer them to a planar mirror to which they adhere by van der Waals forces\,\cite{Riedel2014} [Fig.\,\ref{fig:schematic}(e)]. Individual NV centers in the membranes have PLE linewidths of $\sim 1$\,GHz [Fig.\,\ref{fig:schematic}(b)] increased above the linewidths in the starting material, but still much lower than typical linewidths in diamond nanocrystals. Notably, these linewidths are smaller than the ground state spin triplet splitting of 2.87\,GHz, an essential feature for quantum information applications\,\cite{Faraon2012}.

The miniaturized Fabry-P\'{e}rot cavity consists of a plane bottom mirror and a concave top mirror with radius of curvature $R=16$\,$\mu$m [Fig. \ref{fig:schematic}(a)]. The curved top mirror is fabricated by creating a concave depression in a silica substrate with laser ablation followed by mirror coating\,\cite{Hunger2012,Barbour2011,Greuter2014}. Both bottom and top mirrors are distributed Bragg reflectors (DBRs) with reflectivity $>99.99\,\%$. The bare cavity has a finesse $F \gtrsim 10\,000$. The microcavity can be tuned {\em in situ} with a set of three-axis nanopositioners [Fig.\,\ref{fig:schematic}(e)]. Additionally, the entire microcavity can be moved {\em in situ} with respect to a fixed objective lens, which allows for optimizing mode-matching between the external excitation/detection mode and the microcavity mode\,\cite{Barbour2011,Greuter2014,Greuter2015}. The compact cavity design facilitates low temperature experiments in a liquid helium bath cryostat.

\begin{figure}[t]
\includegraphics[width=0.48\textwidth]{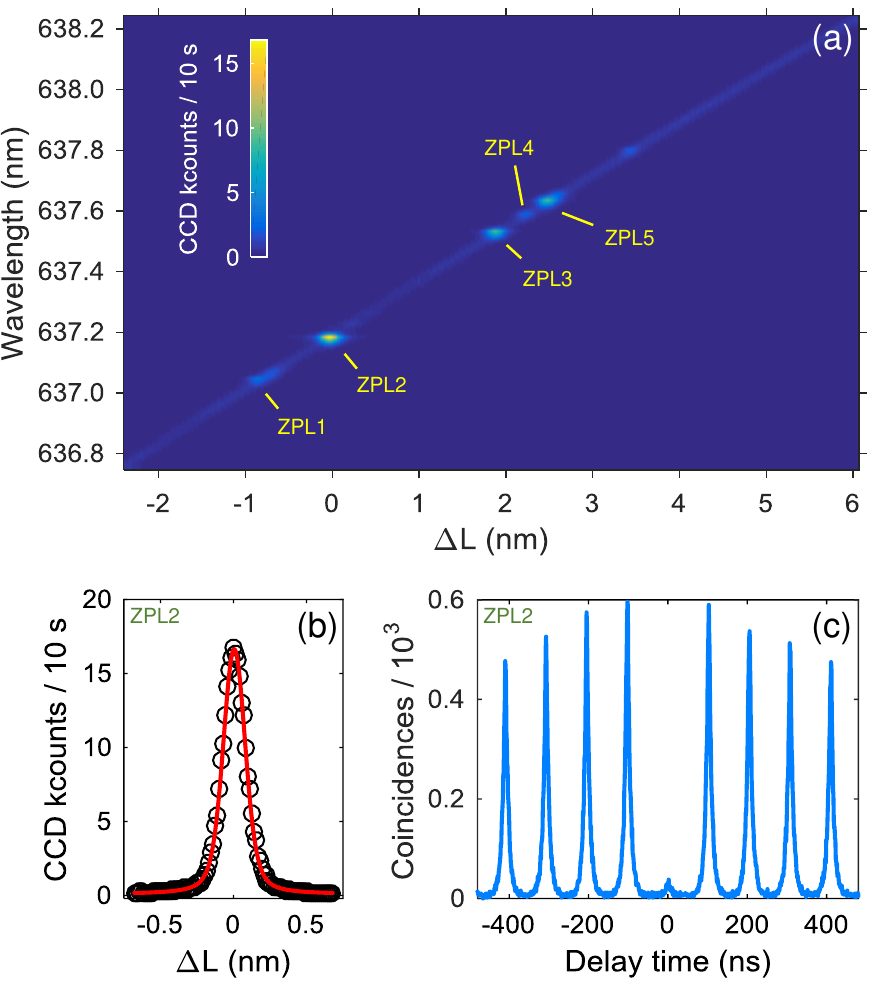}
\caption{(a) Photoluminescence (PL) spectra around the zero phonon line (ZPL) transition for different air-gap detunings $\Delta L$ about $L=1.96\,\mu$m. Each resonance corresponds to the ZPL emission of a single NV center, as labelled. The PL is excited using a pulsed supercontinuum laser source
($P \sim 10\,$mW, $\lambda=560\pm20$\,nm) and detected by a grating spectrometer. (b) Integrated PL versus $L$ for ZPL2 along with a Voigt fit (FWHM Lorentzian contribution $\Gamma_{\rm L}=60.6$\,pm). Together with the cavity dispersion in (a), $\Gamma_{\rm L}$ determines the cavity $Q$-factor, $Q=\lambda/\Gamma_{\rm \lambda}=\lambda/(\Gamma_{\rm L} \cdot {\rm {\rm d}} \lambda/{\rm d}L$) = 58\,500. (c) Photon autocorrelation ($g^{(2)}(t)$) measurement on ZPL2 exhibiting a clear single photon emission $g^{(2)}(0)=0.27$. The ZPL emission is filtered ($637\pm7$\,nm) and analyzed via a Hanbury Brown-Twiss setup (integration time: 45\,000\,s).}
\label{fig:singleNV}
\end{figure}

Figure\,\ref{fig:singleNV}(a) shows photoluminescence (PL) from the diamond membrane-microcavity while detuning the width of the air-gap $L$. The spectra are recorded for the lowest attainable fundamental microcavity mode that comes into resonance with the different ZPL transitions. Notably here, the two orthogonal cavity polarizations are degenerate, which allows for full control over the light polarization. The mirrors are almost in physical contact such that $L$ is dominated by the depth of the curved top mirror ($\sim1\,\mu$m). Spectra were recorded on detuning the microcavity by changing the membrane--top mirror separation, and hence $L$. Weak PL is observed at all $L$ and arises from broadband emission from the diamond membrane. It allows the $L$-dependence of the microcavity mode to be characterized [Fig.\,\ref{fig:singleNV}(a)].

In addition to the weak broadband emission displayed in Fig.\,\ref{fig:singleNV}(a), there are sharp features at specific $L$ which we assign to individual ZPL transitions. The PL from ZPL2 at $L=1.96\,\mu$m is shown for different air-gap detunings $\Delta L$ in Fig.\,\ref{fig:singleNV}(b).
We fit a Voigt profile to the resonance (the Gaussian component accounts for the low-frequency acoustic noise). The FWHM Lorentzian contribution of $\Gamma_{\rm L}=60.6$\,pm yields a finesse of $F=5\,260$. We determine the $Q$-factor of the cavity according to ${Q=\lambda/\Gamma_{\rm \lambda}=\lambda/(\Gamma_{\rm L} \cdot {\rm d} \lambda/{\rm d}L)=2F /( {\rm d} \lambda/{\rm d}L)}$ = 58\,500 with ${\rm d} \lambda/{\rm d}L=0.18$. For fixed $L$, for instance at ZPL2, the microcavity linewidth is $\Gamma_{\rm f}=8.0$\,GHz.

We confirm that the observed resonances are associated with single quantum emitters by performing a photon autocorrelation ($g^{(2)}(t)$) measurement with a Hanbury Brown-Twiss setup. The results on ZPL2 are shown in Fig.\,\ref{fig:singleNV}(c). The small peak at zero delay is a clear signature of single photon emission. The data are analyzed quantitatively by calculating the pulse area of each peak and normalizing the data with the peak area at long delay times (500\,$\mu$s). This gives $g^{(2)}(0)=0.27$, comfortably less than 0.5. Away from $t=0$, $g^{(2)}(t)$ is initially larger than one, a bunching behavior, with a decay as $t$ increases, signifying telegraph noise between a bright and a dark state, as typically observed for an NV center\,\cite{Kurtsiefer2000}.

\begin{figure}[t]
\includegraphics[width=0.48\textwidth]{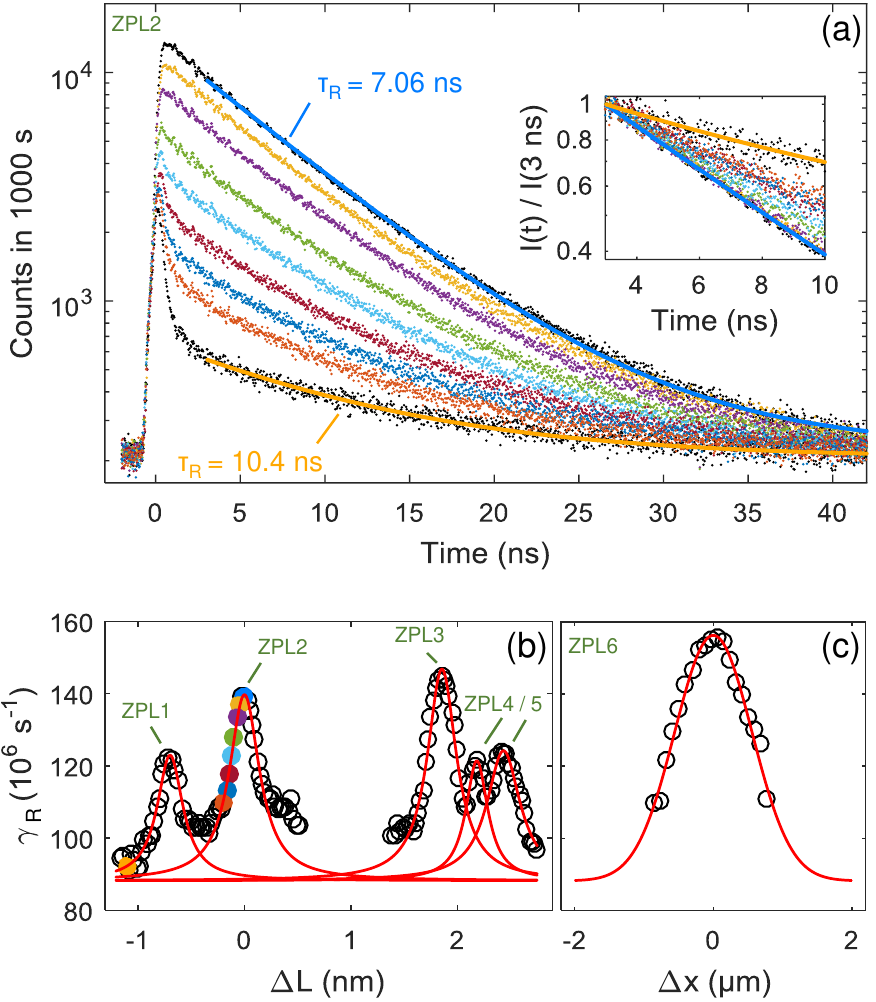}
\caption{(a) PL decay curves of ZPL2 following pulsed excitation as a function of cavity length detuning $\Delta L$ for an acquisition time of 1\,000\,s. The data for delays larger than 3\,ns are fitted to a single-exponential convoluted with the instrumental response. The inset shows the normalized decay curves highlighting the clear change of the decay rate with changing $\Delta L$. (b) Recombination rate $\gamma_{\rm R}$ versus $\Delta L$ for fixed lateral position. For each ZPL resonance, $\gamma_{\rm R}$ exhibits a Purcell effect. The experimental data are fitted to Lorentzian curves with FWHM $\Gamma_{\rm L}=(0.32\pm0.05)$\,nm. The color of the symbols match the decay curves in (a). (c) Recombination rate versus lateral position detuning $\Delta x$ on ZPL6 for zero spectral detuning. The experimental data are fitted to a Gaussian with FWHM 0.80\,$\mu$m.}
\label{fig:lifetime}
\end{figure}

The spectral analysis shows that the microcavity acts as a narrow spectral filter. It is crucial however to demonstrate that the microcavity modifies the behavior of the emitter itself. The key parameter in the weak coupling regime is the decay rate. We therefore turn to excited state lifetime measurements of individual NV centers and tune both the resonance frequency (via an {\em in situ} change of $L$) and the lateral position of the cavity anti-node (via an {\em in situ} change in $(x,y)$). Figure\,\ref{fig:lifetime}(a) shows decay curves for ZPL2 for several different detunings $\Delta L$. There is a clear change in lifetime. To extract the lifetime quantitatively, we note that for large detunings $\Delta L$ the weak background emission represents a significant part of the signal. However, the background emission decays very rapidly: the slower decay process arises from the decay of the NV center. The data for delays larger than 3\,ns are fitted to a single-exponential convoluted with the instrumental response\,\cite{Faraon2011}. The lifetime decreases on detuning $\Delta L$ from 10.4\,ns to 7.06\,ns. This is clear evidence of a Purcell effect.

This assertion is backed up by recording the decay time over a larger range of $\Delta L$ [Fig.\,\ref{fig:lifetime}(b)]. Five ZPLs come into resonance with the microcavity at different values of $\Delta L$ and each shows a clear Purcell effect, an enhanced decay rate $\gamma_{\rm R}$ when in spectral resonance with the microcavity. Each resonance is well described with a Lorentzian function of $\Delta L$. Figure\,\ref{fig:lifetime}(c) shows the results of the alternative experiment, detuning in lateral position $\Delta x$ while maintaining the spectral resonance ($\Delta L=0$), on ZPL6, an NV located at a different location in the diamond membrane. These results also show a Purcell effect, a resonance in $\gamma_{\rm R}$ as a function of $\Delta x$. The $\gamma_{\rm R}$ versus $\Delta x$ data are well fitted by a Gaussian function with FWHM $\Gamma_{\rm x}=0.80\,\mu$m, which represents the lateral extent of the mode in the microcavity. ZPL6 exhibits the largest decay rate on resonance, $\gamma_{\rm R}^{\rm on}=158\cdot10^6$\,s$^{-1}$: this suggests that this NV is positioned close to the optimal depth in the diamond membrane.

We interpret the results in terms of a Purcell enhancement factor $F_{\rm P}$. Without the top mirror the decay time is consistently $\tau_{\rm R}^{0}=12.6$\,ns for all NVs corresponding to a decay rate $\gamma_{\rm R}^{0}=79.4\cdot10^6$\,s$^{-1}$. In terms of the total decay rates, $F_{\rm P}=\gamma_{\rm R}^{\rm on}/\gamma_{\rm R}^{0}=2.0$. We note that, in the microcavity, the decay rate in the limit of large $\Delta L$-detunings is slightly larger than the bulk decay rate [Fig.\,\ref{fig:lifetime}(b)]. We attribute this to non-ZPL emission which is Purcell-enhanced not just at the fundamental microcavity mode but also at higher modes. 

This modest Purcell-enhancement of the total decay rate, $F_{\rm P}=2.0$, masks the large changes in the ZPL emission rate. When the microcavity is tuned into resonance with the ZPL line, the ``on" state, the total decay rate is $\gamma_{\rm R}^{\rm on}=\gamma_{0}^{\rm on}+\gamma_{1}^{\rm on}$ where $\gamma_{0}^{\rm on}$ is the ZPL emission rate; $\gamma_{1}^{\rm on}$ the non-ZPL decay rate. When the microcavity is tuned far out of resonance with the ZPL line, the ``off" state, $\gamma_{\rm R}^{\rm off}=\gamma_{0}^{\rm off}+\gamma_{1}^{\rm off}$. The experiment determines $\gamma_{\rm R}^{\rm on}$ and, by fitting $\gamma_{\rm R}$ as a function of either $\Delta L$ or $\Delta x$ [Fig.\,\ref{fig:lifetime}], $\gamma_{\rm R}^{\rm off}=88.2\cdot10^6$\,s$^{-1}$. To proceed, we note, first, that the non-ZPL emission is so broadband that the small frequency shift of the microcavity mode between the ``on" and ``off" states makes no change to the non-ZPL decay rate, i.e.\ $\gamma_{1}^{\rm on}=\gamma_{1}^{\rm off}$. Secondly, in the ``off" state, ZPL emission takes place predominantly into laterally-propagating modes. This is also the case in a bare membrane such that $\gamma_{0}^{\rm off} \approx \gamma_{0}$ where $\gamma_{0}$ is the ZPL emission rate in the bare membrane. This allows us to determine the Purcell factor for the ZPL alone: $F_{\rm P}^{\rm ZPL}=\gamma_{0}^{\rm on}/\gamma_{0}=(\gamma_{\rm R}^{\rm on}-\gamma_{\rm R}^{\rm off}+\gamma_{0})/\gamma_{0}$. Taking $\gamma_{0}$ as 2.4\,\%\,(5\,\%) of $\gamma_{\rm R}^{0}$, the range of reported NV Debye-Waller factors\,\cite{Faraon2011}, leads to: $\gamma_0=1.91\cdot10^6$\,s$^{-1}$ ($3.97\cdot10^6$\,s$^{-1}$) and hence $F_{\rm P}^{\rm ZPL}=37.7$\,(18.6) (using the experimentally determined values for ZPL6: $\gamma_{\rm R}^{\rm off}=88.2\cdot10^6$\,s$^{-1}$, $\gamma_{\rm R}^{\rm on}=158\cdot10^6$\,s$^{-1}$). The fraction of photons emitted into the ZPL in the ``on" state is $\eta_{\rm ZPL}=F_{\rm P}^{\rm ZPL}\cdot\gamma_{0}/\gamma_{\rm R}^{\rm on}$. For ZPL6, $\eta_{\rm ZPL}=45.4\,\%$\,($46.7\,\%$). We note that $\eta_{\rm ZPL}$, a more important parameter for quantum photonics than $F_{\rm P}^{\rm ZPL}$, depends very weakly on $\gamma_{0}$ which is not known precisely.

We attempt to account quantitatively for the experimental value of $F_{\rm P}^{\rm ZPL}$. The key parameter is the vacuum electric field in the microcavity at the location of the NV center. The vacuum field is a sensitive function of the microcavity geometry which in turn determines the mode structure. In particular, the separation of the fundamental microcavity modes depends on the width of the air-gap; and the frequency separation between the fundamental modes and the higher order modes depends on the radius of curvature of the top mirror. In addition, the mode structure on making large changes to $L$ (which can be described as an anticrossing between air-confined and diamond-confined modes\,\cite{Janitz2015}) depends sensitively on the diamond membrane thickness.

\begin{figure}[t]
\includegraphics[width=0.48\textwidth]{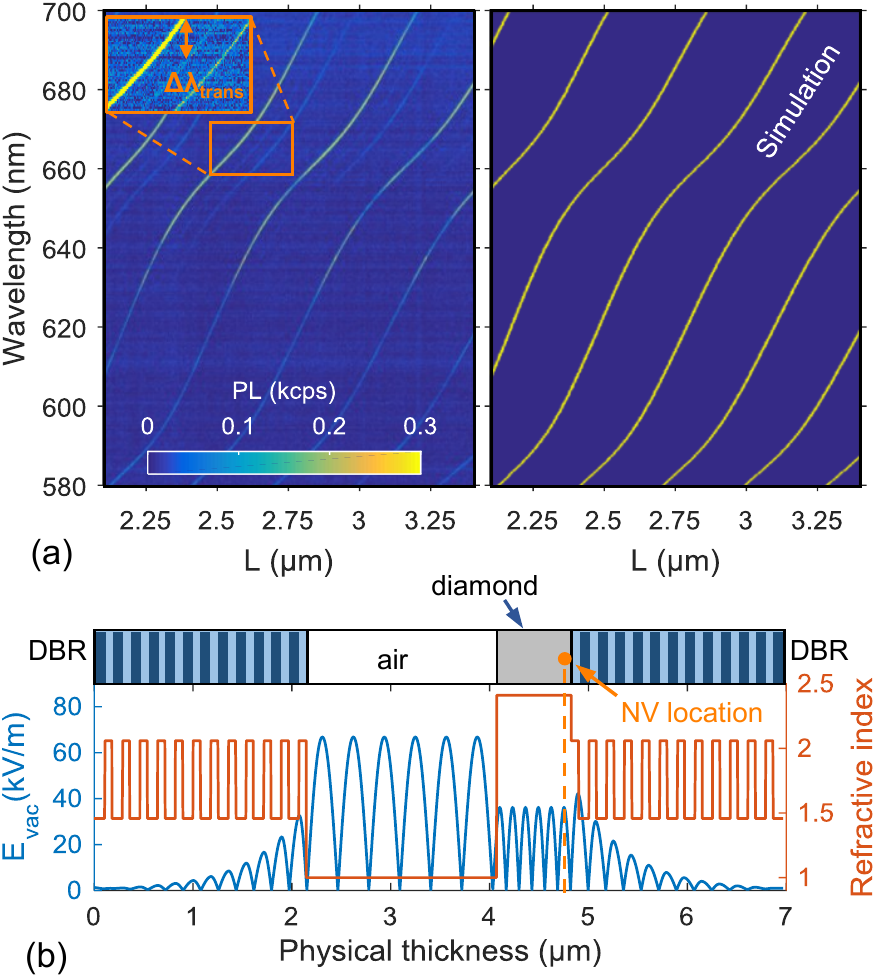}
\caption{(a) Left: measured PL spectra on tuning the microcavity length $L$ over a wide range. The inset highlights the higher order lateral modes. Right: calculated mode dispersion. (b) The layer structure of the microcavity along with the refractive index dependence on $z$. The vacuum electric field $E_{\rm vac}$ is plotted against $z$ for $(x,y)=(0,0)$ for the lowest attainable fundamental microcavity mode. Parameters: diamond thickness $t_{\rm d}=0.77\,\mu$m, air-gap thickness $L=1.96$\,$\mu$m, diamond refractive index $n_{\rm d}=2.41$, refractive indices of Bragg mirror $(n_{\rm H},n_{\rm L})=(2.06,1.46)$, radius of curvature of top mirror $R=16$\,$\mu$m.}
\label{fig:modestructure}
\end{figure}

To measure the mode structure, we excite the diamond--microcavity with a high power of green (560\,nm) light and use the weak broadband emission from the diamond as an internal light source. Figure\,\ref{fig:modestructure}(a) shows the microcavity resonances as a function of wavelength and cavity length spanning several free spectral ranges. The fundamental microcavity modes along with the associated higher order modes are clearly observed [Fig.\,\ref{fig:modestructure}(a)]. We calculate the mode structure by describing the longitudinal standing waves with a transfer-matrix calculation and the longitudinal confinement with Gaussian optics. We achieve excellent agreement with the experiment with $R=16$\,$\mu$m (matching the physical curvature of the mirror), air-gap thickness $L=1.96$\,$\mu$m and $t_{\rm d}=0.77\,\mu$m (corresponding closely to the physical thickness determined with a scanning confocal microscope) and a FWHM beam waist of $0.83\,\mu$m (corresponding to the value determined {\em in situ} on ZPL6) [Fig.\,\ref{fig:modestructure}(b)]. This leads to a maximum vacuum field in the diamond of $E_{\rm vac}=36.2$\,kV/m.

The measurement of $\gamma_{\rm R}^{0}$ enables the optical dipole moment of the NV center to be determined, $d_{\rm NV}/e=0.108$\,nm. Assuming the optical dipole of the ZPL to be parallel to the diamond surface and thus coupling maximally to the two degenerate orthogonal cavity modes, we determine a coupling rate of the ZPL to the vacuum field of $g=d_{\rm NV}\cdot E_{\rm vac}=5.97\cdot10^9$\,s$^{-1}$. Here, we assume a unity internal quantum efficiency of the NV in accordance with a recent extrapolation of measured quantum efficiencies for shallow NVs into the bulk\,\cite{Radko2016}. The measured $Q$-factor determines the photon decay rate out of the microcavity, $\kappa = 2 \pi \Gamma_{\rm f}=5.06\cdot 10^{10}$\,s$^{-1}$. The resultant Purcell factor is $F_{\rm P}^{\rm ZPL}=4g^{2}/\kappa \gamma_{\rm R}^{0}=35.5$, close to the value determined in the experiment. We note, however, that $F_{\rm P}^{\rm ZPL}$ is associated with a systematic error due to the uncertainty of the Debye-Waller factor, the fraction of photons emitted into the ZPL. In fact, the argument can be turned around. The properties of the microcavity are so well understood [Fig.\,\ref{fig:modestructure}(a,b)] that a measurement of $F_{\rm P}^{\rm ZPL}$ constitutes a measurement of $\gamma_{0}/\gamma_{\rm R}^{0}$, the Debye-Waller factor. We find $\gamma_{0}/\gamma_{\rm R}^{0}=2.55\,\%$, which lies within the range of previous estimates\,\cite{Faraon2011}.\\

The ultimate goal is to increase $\eta_{\rm ZPL}$ towards 100\,\% and to collect as many of the ZPL photons as possible. In the present experiment, $\eta_{\rm ZPL}$ is increased to $\sim 46$\,\%. The photon flux is limited by losses (by absorption or scattering) in the dielectric mirror. This loss can be eliminated with better mirrors: 10\,ppm loss dielectric mirrors are available. The present design [Fig.\,\ref{fig:modestructure}(b)] results in a node in the vacuum electric field at the diamond--vacuum interface in order to minimize the scattering losses. The calculations predict that by decreasing the diamond membrane thickness $t_{\rm d}$ and employing state-of-the-art laser ablation techniques for fabricating shallow depressions with a smaller radius of curvature (depth $\sim 400\,$nm, $R \sim 5.5\,\mu$m\,\cite{Najer2017}) significantly boosts the maximum vacuum field $E_{\rm vac}$ in the diamond. For instance, decreasing $t_{\rm d}$ from $0.77\,\mu$m to $3\lambda_\text{ZPL}/4=198\,$nm in combination with the aforementioned improved microcavity design, allowing for $L=3\lambda_\text{ZPL}/4=478$\,nm, increases the maximum vacuum field from $E_{\rm vac}=36.2$\,kV/m to $E_{\rm vac}=85.7$\,kV/m. We estimate that photon collection is maximized for $\kappa=2g$, which requires a $Q$-factor of only 128\,000; the resulting Purcell factor and ZPL emission probability are $F_{\rm P}^{\rm ZPL}=356$ and $\eta_{\rm ZPL}=87.9\,\%$, respectively. If the diamond surface can be made sufficiently smooth then it becomes possible to change the design such that there is an anti-node at the diamond--vacuum interface. This boosts the vacuum electric field at the anti-node in the diamond membrane: the global maximum of the vacuum field is now located in the diamond rather than in the air-gap\,\cite{Barbour2011}. For instance, for $t_{\rm d}=\lambda_\text{ZPL}/2=132$\,nm, $L=\lambda_\text{ZPL}=637$\,nm, $E_{\rm vac}=127$\,kV/m and, again choosing $\kappa=2g$ (requiring $Q=86\,500$), $F_{\rm P}^{\rm ZPL}=527$, and $\eta_{\rm ZPL}=91.5\,\%$. In addition to providing a massive boost to the ZPL fraction a further advantage of a large Purcell enhancement is that the NV transform limit increases, for $F_{\rm P}^{\rm ZPL}=527 (356)$ to $\Gamma_{\rm f}=182$\,MHz (127\,MHz). Implementing these improvements would therefore mitigate the constraints on the spectral stability of the NV ZPL transition. Only a slight improvement on the linewidths reported here is required in order to create a high fraction of indistinguishable photons. In combination with an estimated twofold enhancement in the collection efficiency the spin-spin entanglement rate would be boosted by more than six orders of magnitude\,\cite{Bogdanovic2017}. 

Based on the work present here, we propose that a miniaturized Fabry-P\'{e}rot microcavity using a thin diamond membrane is a very attractive platform for quantum technology applications. Furthermore, the miniaturized Fabry-P\'{e}rot microcavity has a rather generic design such that cavity enhancement is an immediate possibility also with other color centers in for instance diamond\,\cite{Aharonovich2014, Siyushev2016, Bhaskar2016} and SiC\,\cite{Koehl2011,Castelletto2014,Riedel2012}. 

We acknowledge financial support from NCCR QSIT, a competence center funded by SNF; the Swiss Nanoscience Institute (SNI); the EU FP7 project DIADEMS (grant No.\,611143); ITN networks S$^{3}$NANO and SpinNANO; and through SNF Grants 200021\_143697 and 200021\_169321. 

\bibliographystyle{apsrev4-1}

\begin{thebibliography}{43}%
\makeatletter
\providecommand \@ifxundefined [1]{%
 \@ifx{#1\undefined}
}%
\providecommand \@ifnum [1]{%
 \ifnum #1\expandafter \@firstoftwo
 \else \expandafter \@secondoftwo
 \fi
}%
\providecommand \@ifx [1]{%
 \ifx #1\expandafter \@firstoftwo
 \else \expandafter \@secondoftwo
 \fi
}%
\providecommand \natexlab [1]{#1}%
\providecommand \enquote  [1]{``#1''}%
\providecommand \bibnamefont  [1]{#1}%
\providecommand \bibfnamefont [1]{#1}%
\providecommand \citenamefont [1]{#1}%
\providecommand \href@noop [0]{\@secondoftwo}%
\providecommand \href [0]{\begingroup \@sanitize@url \@href}%
\providecommand \@href[1]{\@@startlink{#1}\@@href}%
\providecommand \@@href[1]{\endgroup#1\@@endlink}%
\providecommand \@sanitize@url [0]{\catcode `\\12\catcode `\$12\catcode
  `\&12\catcode `\#12\catcode `\^12\catcode `\_12\catcode `\%12\relax}%
\providecommand \@@startlink[1]{}%
\providecommand \@@endlink[0]{}%
\providecommand \url  [0]{\begingroup\@sanitize@url \@url }%
\providecommand \@url [1]{\endgroup\@href {#1}{\urlprefix }}%
\providecommand \urlprefix  [0]{URL }%
\providecommand \Eprint [0]{\href }%
\providecommand \doibase [0]{http://dx.doi.org/}%
\providecommand \selectlanguage [0]{\@gobble}%
\providecommand \bibinfo  [0]{\@secondoftwo}%
\providecommand \bibfield  [0]{\@secondoftwo}%
\providecommand \translation [1]{[#1]}%
\providecommand \BibitemOpen [0]{}%
\providecommand \bibitemStop [0]{}%
\providecommand \bibitemNoStop [0]{.\EOS\space}%
\providecommand \EOS [0]{\spacefactor3000\relax}%
\providecommand \BibitemShut  [1]{\csname bibitem#1\endcsname}%
\let\auto@bib@innerbib\@empty
\bibitem [{\citenamefont {Jelezko}\ \emph {et~al.}(2004)\citenamefont
  {Jelezko}, \citenamefont {Gaebel}, \citenamefont {Popa}, \citenamefont
  {Gruber},\ and\ \citenamefont {Wrachtrup}}]{Jelezko2004}%
  \BibitemOpen
  \bibfield  {author} {\bibinfo {author} {\bibfnamefont {F.}~\bibnamefont
  {Jelezko}}, \bibinfo {author} {\bibfnamefont {T.}~\bibnamefont {Gaebel}},
  \bibinfo {author} {\bibfnamefont {I.}~\bibnamefont {Popa}}, \bibinfo {author}
  {\bibfnamefont {A.}~\bibnamefont {Gruber}}, \ and\ \bibinfo {author}
  {\bibfnamefont {J.}~\bibnamefont {Wrachtrup}},\ }\href {\doibase
  10.1103/PhysRevLett.92.076401} {\bibfield  {journal} {\bibinfo  {journal}
  {Phys. Rev. Lett.}\ }\textbf {\bibinfo {volume} {92}},\ \bibinfo {pages}
  {076401} (\bibinfo {year} {2004})}\BibitemShut {NoStop}%
\bibitem [{\citenamefont {Balasubramanian}\ \emph {et~al.}(2009)\citenamefont
  {Balasubramanian}, \citenamefont {Neumann}, \citenamefont {Twitchen},
  \citenamefont {Markham}, \citenamefont {Kolesov}, \citenamefont {Mizuochi},
  \citenamefont {Isoya}, \citenamefont {Achard}, \citenamefont {Beck},
  \citenamefont {Tissler}, \citenamefont {Jacques}, \citenamefont {Hemmer},
  \citenamefont {Jelezko},\ and\ \citenamefont
  {Wrachtrup}}]{Balasubramanian2009}%
  \BibitemOpen
  \bibfield  {author} {\bibinfo {author} {\bibfnamefont {G.}~\bibnamefont
  {Balasubramanian}}, \bibinfo {author} {\bibfnamefont {P.}~\bibnamefont
  {Neumann}}, \bibinfo {author} {\bibfnamefont {D.}~\bibnamefont {Twitchen}},
  \bibinfo {author} {\bibfnamefont {M.}~\bibnamefont {Markham}}, \bibinfo
  {author} {\bibfnamefont {R.}~\bibnamefont {Kolesov}}, \bibinfo {author}
  {\bibfnamefont {N.}~\bibnamefont {Mizuochi}}, \bibinfo {author}
  {\bibfnamefont {J.}~\bibnamefont {Isoya}}, \bibinfo {author} {\bibfnamefont
  {J.}~\bibnamefont {Achard}}, \bibinfo {author} {\bibfnamefont
  {J.}~\bibnamefont {Beck}}, \bibinfo {author} {\bibfnamefont {J.}~\bibnamefont
  {Tissler}}, \bibinfo {author} {\bibfnamefont {V.}~\bibnamefont {Jacques}},
  \bibinfo {author} {\bibfnamefont {P.~R.}\ \bibnamefont {Hemmer}}, \bibinfo
  {author} {\bibfnamefont {F.}~\bibnamefont {Jelezko}}, \ and\ \bibinfo
  {author} {\bibfnamefont {J.}~\bibnamefont {Wrachtrup}},\ }\href {\doibase
  10.1038/nmat2420} {\bibfield  {journal} {\bibinfo  {journal} {Nat. Mater.}\
  }\textbf {\bibinfo {volume} {8}},\ \bibinfo {pages} {383} (\bibinfo {year}
  {2009})}\BibitemShut {NoStop}%
\bibitem [{\citenamefont {Kurtsiefer}\ \emph {et~al.}(2000)\citenamefont
  {Kurtsiefer}, \citenamefont {Mayer}, \citenamefont {Zarda},\ and\
  \citenamefont {Weinfurter}}]{Kurtsiefer2000}%
  \BibitemOpen
  \bibfield  {author} {\bibinfo {author} {\bibfnamefont {C.}~\bibnamefont
  {Kurtsiefer}}, \bibinfo {author} {\bibfnamefont {S.}~\bibnamefont {Mayer}},
  \bibinfo {author} {\bibfnamefont {P.}~\bibnamefont {Zarda}}, \ and\ \bibinfo
  {author} {\bibfnamefont {H.}~\bibnamefont {Weinfurter}},\ }\href {\doibase
  10.1103/PhysRevLett.85.290} {\bibfield  {journal} {\bibinfo  {journal} {Phys.
  Rev. Lett.}\ }\textbf {\bibinfo {volume} {85}},\ \bibinfo {pages} {290}
  (\bibinfo {year} {2000})}\BibitemShut {NoStop}%
\bibitem [{\citenamefont {Dutt}\ \emph {et~al.}(2007)\citenamefont {Dutt},
  \citenamefont {Childress}, \citenamefont {Jiang}, \citenamefont {Togan},
  \citenamefont {Maze}, \citenamefont {Jelezko}, \citenamefont {Zibrov},
  \citenamefont {Hemmer},\ and\ \citenamefont {Lukin}}]{Dutt2007}%
  \BibitemOpen
  \bibfield  {author} {\bibinfo {author} {\bibfnamefont {M.~V.~G.}\
  \bibnamefont {Dutt}}, \bibinfo {author} {\bibfnamefont {L.}~\bibnamefont
  {Childress}}, \bibinfo {author} {\bibfnamefont {L.}~\bibnamefont {Jiang}},
  \bibinfo {author} {\bibfnamefont {E.}~\bibnamefont {Togan}}, \bibinfo
  {author} {\bibfnamefont {J.}~\bibnamefont {Maze}}, \bibinfo {author}
  {\bibfnamefont {F.}~\bibnamefont {Jelezko}}, \bibinfo {author} {\bibfnamefont
  {A.~S.}\ \bibnamefont {Zibrov}}, \bibinfo {author} {\bibfnamefont {P.~R.}\
  \bibnamefont {Hemmer}}, \ and\ \bibinfo {author} {\bibfnamefont {M.~D.}\
  \bibnamefont {Lukin}},\ }\href
  {http://science.sciencemag.org/content/316/5829/1312} {\bibfield  {journal}
  {\bibinfo  {journal} {Science}\ }\textbf {\bibinfo {volume} {316}} (\bibinfo
  {year} {2007})}\BibitemShut {NoStop}%
\bibitem [{\citenamefont {Togan}\ \emph {et~al.}(2010)\citenamefont {Togan},
  \citenamefont {Chu}, \citenamefont {Trifonov}, \citenamefont {Jiang},
  \citenamefont {Maze}, \citenamefont {Childress}, \citenamefont {Dutt},
  \citenamefont {S{\o}rensen}, \citenamefont {Hemmer}, \citenamefont {Zibrov},\
  and\ \citenamefont {Lukin}}]{Togan2010}%
  \BibitemOpen
  \bibfield  {author} {\bibinfo {author} {\bibfnamefont {E.}~\bibnamefont
  {Togan}}, \bibinfo {author} {\bibfnamefont {Y.}~\bibnamefont {Chu}}, \bibinfo
  {author} {\bibfnamefont {A.~S.}\ \bibnamefont {Trifonov}}, \bibinfo {author}
  {\bibfnamefont {L.}~\bibnamefont {Jiang}}, \bibinfo {author} {\bibfnamefont
  {J.}~\bibnamefont {Maze}}, \bibinfo {author} {\bibfnamefont {L.}~\bibnamefont
  {Childress}}, \bibinfo {author} {\bibfnamefont {M.~V.~G.}\ \bibnamefont
  {Dutt}}, \bibinfo {author} {\bibfnamefont {A.~S.}\ \bibnamefont
  {S{\o}rensen}}, \bibinfo {author} {\bibfnamefont {P.~R.}\ \bibnamefont
  {Hemmer}}, \bibinfo {author} {\bibfnamefont {A.~S.}\ \bibnamefont {Zibrov}},
  \ and\ \bibinfo {author} {\bibfnamefont {M.~D.}\ \bibnamefont {Lukin}},\
  }\href {\doibase 10.1038/nature09256} {\bibfield  {journal} {\bibinfo
  {journal} {Nature}\ }\textbf {\bibinfo {volume} {466}},\ \bibinfo {pages}
  {730} (\bibinfo {year} {2010})}\BibitemShut {NoStop}%
\bibitem [{\citenamefont {Bernien}\ \emph {et~al.}(2012)\citenamefont
  {Bernien}, \citenamefont {Childress}, \citenamefont {Robledo}, \citenamefont
  {Markham}, \citenamefont {Twitchen},\ and\ \citenamefont
  {Hanson}}]{Bernien2012}%
  \BibitemOpen
  \bibfield  {author} {\bibinfo {author} {\bibfnamefont {H.}~\bibnamefont
  {Bernien}}, \bibinfo {author} {\bibfnamefont {L.}~\bibnamefont {Childress}},
  \bibinfo {author} {\bibfnamefont {L.}~\bibnamefont {Robledo}}, \bibinfo
  {author} {\bibfnamefont {M.}~\bibnamefont {Markham}}, \bibinfo {author}
  {\bibfnamefont {D.}~\bibnamefont {Twitchen}}, \ and\ \bibinfo {author}
  {\bibfnamefont {R.}~\bibnamefont {Hanson}},\ }\href {\doibase
  10.1103/PhysRevLett.108.043604} {\bibfield  {journal} {\bibinfo  {journal}
  {Phys. Rev. Lett.}\ }\textbf {\bibinfo {volume} {108}},\ \bibinfo {pages}
  {043604} (\bibinfo {year} {2012})}\BibitemShut {NoStop}%
\bibitem [{\citenamefont {Sipahigil}\ \emph {et~al.}(2012)\citenamefont
  {Sipahigil}, \citenamefont {Goldman}, \citenamefont {Togan}, \citenamefont
  {Chu}, \citenamefont {Markham}, \citenamefont {Twitchen}, \citenamefont
  {Zibrov}, \citenamefont {Kubanek},\ and\ \citenamefont
  {Lukin}}]{Sipahigil2012}%
  \BibitemOpen
  \bibfield  {author} {\bibinfo {author} {\bibfnamefont {A.}~\bibnamefont
  {Sipahigil}}, \bibinfo {author} {\bibfnamefont {M.~L.}\ \bibnamefont
  {Goldman}}, \bibinfo {author} {\bibfnamefont {E.}~\bibnamefont {Togan}},
  \bibinfo {author} {\bibfnamefont {Y.}~\bibnamefont {Chu}}, \bibinfo {author}
  {\bibfnamefont {M.}~\bibnamefont {Markham}}, \bibinfo {author} {\bibfnamefont
  {D.~J.}\ \bibnamefont {Twitchen}}, \bibinfo {author} {\bibfnamefont {A.~S.}\
  \bibnamefont {Zibrov}}, \bibinfo {author} {\bibfnamefont {A.}~\bibnamefont
  {Kubanek}}, \ and\ \bibinfo {author} {\bibfnamefont {M.~D.}\ \bibnamefont
  {Lukin}},\ }\href {\doibase 10.1103/PhysRevLett.108.143601} {\bibfield
  {journal} {\bibinfo  {journal} {Phys. Rev. Lett.}\ }\textbf {\bibinfo
  {volume} {108}},\ \bibinfo {pages} {143601} (\bibinfo {year}
  {2012})}\BibitemShut {NoStop}%
\bibitem [{\citenamefont {Pfaff}\ \emph {et~al.}(2014)\citenamefont {Pfaff},
  \citenamefont {Hensen}, \citenamefont {Bernien}, \citenamefont {van Dam},
  \citenamefont {Blok}, \citenamefont {Taminiau}, \citenamefont {Tiggelman},
  \citenamefont {Schouten}, \citenamefont {Markham}, \citenamefont {Twitchen},\
  and\ \citenamefont {Hanson}}]{Pfaff2014}%
  \BibitemOpen
  \bibfield  {author} {\bibinfo {author} {\bibfnamefont {W.}~\bibnamefont
  {Pfaff}}, \bibinfo {author} {\bibfnamefont {B.}~\bibnamefont {Hensen}},
  \bibinfo {author} {\bibfnamefont {H.}~\bibnamefont {Bernien}}, \bibinfo
  {author} {\bibfnamefont {S.~B.}\ \bibnamefont {van Dam}}, \bibinfo {author}
  {\bibfnamefont {M.~S.}\ \bibnamefont {Blok}}, \bibinfo {author}
  {\bibfnamefont {T.~H.}\ \bibnamefont {Taminiau}}, \bibinfo {author}
  {\bibfnamefont {M.~J.}\ \bibnamefont {Tiggelman}}, \bibinfo {author}
  {\bibfnamefont {R.~N.}\ \bibnamefont {Schouten}}, \bibinfo {author}
  {\bibfnamefont {M.}~\bibnamefont {Markham}}, \bibinfo {author} {\bibfnamefont
  {D.~J.}\ \bibnamefont {Twitchen}}, \ and\ \bibinfo {author} {\bibfnamefont
  {R.}~\bibnamefont {Hanson}},\ }\href {\doibase 10.1126/science.1253512}
  {\bibfield  {journal} {\bibinfo  {journal} {Science}\ }\textbf {\bibinfo
  {volume} {345}},\ \bibinfo {pages} {532} (\bibinfo {year}
  {2014})}\BibitemShut {NoStop}%
\bibitem [{\citenamefont {Hensen}\ \emph {et~al.}(2015)\citenamefont {Hensen},
  \citenamefont {Bernien}, \citenamefont {Dr{\'{e}}au}, \citenamefont
  {Reiserer}, \citenamefont {Kalb}, \citenamefont {Blok}, \citenamefont
  {Ruitenberg}, \citenamefont {Vermeulen}, \citenamefont {Schouten},
  \citenamefont {Abell{\'{a}}n}, \citenamefont {Amaya}, \citenamefont
  {Pruneri}, \citenamefont {Mitchell}, \citenamefont {Markham}, \citenamefont
  {Twitchen}, \citenamefont {Elkouss}, \citenamefont {Wehner}, \citenamefont
  {Taminiau},\ and\ \citenamefont {Hanson}}]{Hensen2015}%
  \BibitemOpen
  \bibfield  {author} {\bibinfo {author} {\bibfnamefont {B.}~\bibnamefont
  {Hensen}}, \bibinfo {author} {\bibfnamefont {H.}~\bibnamefont {Bernien}},
  \bibinfo {author} {\bibfnamefont {A.~E.}\ \bibnamefont {Dr{\'{e}}au}},
  \bibinfo {author} {\bibfnamefont {A.}~\bibnamefont {Reiserer}}, \bibinfo
  {author} {\bibfnamefont {N.}~\bibnamefont {Kalb}}, \bibinfo {author}
  {\bibfnamefont {M.~S.}\ \bibnamefont {Blok}}, \bibinfo {author}
  {\bibfnamefont {J.}~\bibnamefont {Ruitenberg}}, \bibinfo {author}
  {\bibfnamefont {R.~F.~L.}\ \bibnamefont {Vermeulen}}, \bibinfo {author}
  {\bibfnamefont {R.~N.}\ \bibnamefont {Schouten}}, \bibinfo {author}
  {\bibfnamefont {C.}~\bibnamefont {Abell{\'{a}}n}}, \bibinfo {author}
  {\bibfnamefont {W.}~\bibnamefont {Amaya}}, \bibinfo {author} {\bibfnamefont
  {V.}~\bibnamefont {Pruneri}}, \bibinfo {author} {\bibfnamefont {M.~W.}\
  \bibnamefont {Mitchell}}, \bibinfo {author} {\bibfnamefont {M.}~\bibnamefont
  {Markham}}, \bibinfo {author} {\bibfnamefont {D.~J.}\ \bibnamefont
  {Twitchen}}, \bibinfo {author} {\bibfnamefont {D.}~\bibnamefont {Elkouss}},
  \bibinfo {author} {\bibfnamefont {S.}~\bibnamefont {Wehner}}, \bibinfo
  {author} {\bibfnamefont {T.~H.}\ \bibnamefont {Taminiau}}, \ and\ \bibinfo
  {author} {\bibfnamefont {R.}~\bibnamefont {Hanson}},\ }\href {\doibase
  10.1038/nature15759} {\bibfield  {journal} {\bibinfo  {journal} {Nature}\
  }\textbf {\bibinfo {volume} {526}},\ \bibinfo {pages} {682} (\bibinfo {year}
  {2015})}\BibitemShut {NoStop}%
\bibitem [{\citenamefont {Gao}\ \emph {et~al.}(2015)\citenamefont {Gao},
  \citenamefont {Imamoglu}, \citenamefont {Bernien},\ and\ \citenamefont
  {Hanson}}]{Gao2015}%
  \BibitemOpen
  \bibfield  {author} {\bibinfo {author} {\bibfnamefont {W.~B.}\ \bibnamefont
  {Gao}}, \bibinfo {author} {\bibfnamefont {A.}~\bibnamefont {Imamoglu}},
  \bibinfo {author} {\bibfnamefont {H.}~\bibnamefont {Bernien}}, \ and\
  \bibinfo {author} {\bibfnamefont {R.}~\bibnamefont {Hanson}},\ }\href
  {\doibase 10.1038/nphoton.2015.58} {\bibfield  {journal} {\bibinfo  {journal}
  {Nat. Photonics}\ }\textbf {\bibinfo {volume} {9}},\ \bibinfo {pages} {363}
  (\bibinfo {year} {2015})}\BibitemShut {NoStop}%
\bibitem [{\citenamefont {Barclay}\ \emph {et~al.}(2011)\citenamefont
  {Barclay}, \citenamefont {Fu}, \citenamefont {Santori}, \citenamefont
  {Faraon},\ and\ \citenamefont {Beausoleil}}]{Barclay2011}%
  \BibitemOpen
  \bibfield  {author} {\bibinfo {author} {\bibfnamefont {P.~E.}\ \bibnamefont
  {Barclay}}, \bibinfo {author} {\bibfnamefont {K.-M.~C.}\ \bibnamefont {Fu}},
  \bibinfo {author} {\bibfnamefont {C.}~\bibnamefont {Santori}}, \bibinfo
  {author} {\bibfnamefont {A.}~\bibnamefont {Faraon}}, \ and\ \bibinfo {author}
  {\bibfnamefont {R.~G.}\ \bibnamefont {Beausoleil}},\ }\href {\doibase
  10.1103/PhysRevX.1.011007} {\bibfield  {journal} {\bibinfo  {journal} {Phys.
  Rev. X}\ }\textbf {\bibinfo {volume} {1}},\ \bibinfo {pages} {011007}
  (\bibinfo {year} {2011})}\BibitemShut {NoStop}%
\bibitem [{\citenamefont {Kaupp}\ \emph {et~al.}(2013)\citenamefont {Kaupp},
  \citenamefont {Deutsch}, \citenamefont {Chang}, \citenamefont {Reichel},
  \citenamefont {H{\"{a}}nsch},\ and\ \citenamefont {Hunger}}]{Kaupp2013}%
  \BibitemOpen
  \bibfield  {author} {\bibinfo {author} {\bibfnamefont {H.}~\bibnamefont
  {Kaupp}}, \bibinfo {author} {\bibfnamefont {C.}~\bibnamefont {Deutsch}},
  \bibinfo {author} {\bibfnamefont {H.-C.}\ \bibnamefont {Chang}}, \bibinfo
  {author} {\bibfnamefont {J.}~\bibnamefont {Reichel}}, \bibinfo {author}
  {\bibfnamefont {T.~W.}\ \bibnamefont {H{\"{a}}nsch}}, \ and\ \bibinfo
  {author} {\bibfnamefont {D.}~\bibnamefont {Hunger}},\ }\href {\doibase
  10.1103/PhysRevA.88.053812} {\bibfield  {journal} {\bibinfo  {journal} {Phys.
  Rev. A}\ }\textbf {\bibinfo {volume} {88}},\ \bibinfo {pages} {053812}
  (\bibinfo {year} {2013})}\BibitemShut {NoStop}%
\bibitem [{\citenamefont {Khitrova}\ \emph {et~al.}(2006)\citenamefont
  {Khitrova}, \citenamefont {Gibbs}, \citenamefont {Kira}, \citenamefont
  {Koch},\ and\ \citenamefont {Scherer}}]{Khitrova2006}%
  \BibitemOpen
  \bibfield  {author} {\bibinfo {author} {\bibfnamefont {G.}~\bibnamefont
  {Khitrova}}, \bibinfo {author} {\bibfnamefont {H.~M.}\ \bibnamefont {Gibbs}},
  \bibinfo {author} {\bibfnamefont {M.}~\bibnamefont {Kira}}, \bibinfo {author}
  {\bibfnamefont {S.~W.}\ \bibnamefont {Koch}}, \ and\ \bibinfo {author}
  {\bibfnamefont {A.}~\bibnamefont {Scherer}},\ }\href {\doibase
  10.1038/nphys227} {\bibfield  {journal} {\bibinfo  {journal} {Nat. Phys.}\
  }\textbf {\bibinfo {volume} {2}},\ \bibinfo {pages} {81} (\bibinfo {year}
  {2006})}\BibitemShut {NoStop}%
\bibitem [{\citenamefont {Sipahigil}\ \emph {et~al.}(2016)\citenamefont
  {Sipahigil}, \citenamefont {Evans}, \citenamefont {Sukachev}, \citenamefont
  {Burek}, \citenamefont {Borregaard}, \citenamefont {Bhaskar}, \citenamefont
  {Nguyen}, \citenamefont {Pacheco}, \citenamefont {Atikian}, \citenamefont
  {Meuwly}, \citenamefont {Camacho}, \citenamefont {Jelezko}, \citenamefont
  {Bielejec}, \citenamefont {Park}, \citenamefont {Lon{\v{c}}ar},\ and\
  \citenamefont {Lukin}}]{Sipahigil2016}%
  \BibitemOpen
  \bibfield  {author} {\bibinfo {author} {\bibfnamefont {A.}~\bibnamefont
  {Sipahigil}}, \bibinfo {author} {\bibfnamefont {R.~E.}\ \bibnamefont
  {Evans}}, \bibinfo {author} {\bibfnamefont {D.~D.}\ \bibnamefont {Sukachev}},
  \bibinfo {author} {\bibfnamefont {M.~J.}\ \bibnamefont {Burek}}, \bibinfo
  {author} {\bibfnamefont {J.}~\bibnamefont {Borregaard}}, \bibinfo {author}
  {\bibfnamefont {M.~K.}\ \bibnamefont {Bhaskar}}, \bibinfo {author}
  {\bibfnamefont {C.~T.}\ \bibnamefont {Nguyen}}, \bibinfo {author}
  {\bibfnamefont {J.~L.}\ \bibnamefont {Pacheco}}, \bibinfo {author}
  {\bibfnamefont {H.~A.}\ \bibnamefont {Atikian}}, \bibinfo {author}
  {\bibfnamefont {C.}~\bibnamefont {Meuwly}}, \bibinfo {author} {\bibfnamefont
  {R.~M.}\ \bibnamefont {Camacho}}, \bibinfo {author} {\bibfnamefont
  {F.}~\bibnamefont {Jelezko}}, \bibinfo {author} {\bibfnamefont
  {E.}~\bibnamefont {Bielejec}}, \bibinfo {author} {\bibfnamefont
  {H.}~\bibnamefont {Park}}, \bibinfo {author} {\bibfnamefont {M.}~\bibnamefont
  {Lon{\v{c}}ar}}, \ and\ \bibinfo {author} {\bibfnamefont {M.~D.}\
  \bibnamefont {Lukin}},\ }\href {\doibase 10.1126/science.aah6875} {\bibfield
  {journal} {\bibinfo  {journal} {Science}\ }\textbf {\bibinfo {volume}
  {354}},\ \bibinfo {pages} {847} (\bibinfo {year} {2016})}\BibitemShut
  {NoStop}%
\bibitem [{\citenamefont {Riedrich-M{\"{o}}ller}\ \emph
  {et~al.}(2015)\citenamefont {Riedrich-M{\"{o}}ller}, \citenamefont
  {Pezzagna}, \citenamefont {Meijer}, \citenamefont {Pauly}, \citenamefont
  {M{\"{u}}cklich}, \citenamefont {Markham}, \citenamefont {Edmonds},\ and\
  \citenamefont {Becher}}]{Riedrich2015}%
  \BibitemOpen
  \bibfield  {author} {\bibinfo {author} {\bibfnamefont {J.}~\bibnamefont
  {Riedrich-M{\"{o}}ller}}, \bibinfo {author} {\bibfnamefont {S.}~\bibnamefont
  {Pezzagna}}, \bibinfo {author} {\bibfnamefont {J.}~\bibnamefont {Meijer}},
  \bibinfo {author} {\bibfnamefont {C.}~\bibnamefont {Pauly}}, \bibinfo
  {author} {\bibfnamefont {F.}~\bibnamefont {M{\"{u}}cklich}}, \bibinfo
  {author} {\bibfnamefont {M.}~\bibnamefont {Markham}}, \bibinfo {author}
  {\bibfnamefont {A.~M.}\ \bibnamefont {Edmonds}}, \ and\ \bibinfo {author}
  {\bibfnamefont {C.}~\bibnamefont {Becher}},\ }\href {\doibase
  10.1063/1.4922117} {\bibfield  {journal} {\bibinfo  {journal} {Appl. Phys.
  Lett.}\ }\textbf {\bibinfo {volume} {106}},\ \bibinfo {pages} {221103}
  (\bibinfo {year} {2015})}\BibitemShut {NoStop}%
\bibitem [{\citenamefont {Faraon}\ \emph {et~al.}(2012)\citenamefont {Faraon},
  \citenamefont {Santori}, \citenamefont {Huang}, \citenamefont {Acosta},\ and\
  \citenamefont {Beausoleil}}]{Faraon2012}%
  \BibitemOpen
  \bibfield  {author} {\bibinfo {author} {\bibfnamefont {A.}~\bibnamefont
  {Faraon}}, \bibinfo {author} {\bibfnamefont {C.}~\bibnamefont {Santori}},
  \bibinfo {author} {\bibfnamefont {Z.}~\bibnamefont {Huang}}, \bibinfo
  {author} {\bibfnamefont {V.~M.}\ \bibnamefont {Acosta}}, \ and\ \bibinfo
  {author} {\bibfnamefont {R.~G.}\ \bibnamefont {Beausoleil}},\ }\href
  {\doibase 10.1103/PhysRevLett.109.033604} {\bibfield  {journal} {\bibinfo
  {journal} {Phys. Rev. Lett.}\ }\textbf {\bibinfo {volume} {109}},\ \bibinfo
  {pages} {033604} (\bibinfo {year} {2012})}\BibitemShut {NoStop}%
\bibitem [{\citenamefont {Hausmann}\ \emph {et~al.}(2013)\citenamefont
  {Hausmann}, \citenamefont {Shields}, \citenamefont {Quan}, \citenamefont
  {Chu}, \citenamefont {de~Leon}, \citenamefont {Evans}, \citenamefont {Burek},
  \citenamefont {Zibrov}, \citenamefont {Markham}, \citenamefont {Twitchen},
  \citenamefont {Park}, \citenamefont {Lukin},\ and\ \citenamefont
  {Loncar}}]{Hausmann2013}%
  \BibitemOpen
  \bibfield  {author} {\bibinfo {author} {\bibfnamefont {B.~J.~M.}\
  \bibnamefont {Hausmann}}, \bibinfo {author} {\bibfnamefont {B.~J.}\
  \bibnamefont {Shields}}, \bibinfo {author} {\bibfnamefont {Q.}~\bibnamefont
  {Quan}}, \bibinfo {author} {\bibfnamefont {Y.}~\bibnamefont {Chu}}, \bibinfo
  {author} {\bibfnamefont {N.~P.}\ \bibnamefont {de~Leon}}, \bibinfo {author}
  {\bibfnamefont {R.}~\bibnamefont {Evans}}, \bibinfo {author} {\bibfnamefont
  {M.~J.}\ \bibnamefont {Burek}}, \bibinfo {author} {\bibfnamefont {A.~S.}\
  \bibnamefont {Zibrov}}, \bibinfo {author} {\bibfnamefont {M.}~\bibnamefont
  {Markham}}, \bibinfo {author} {\bibfnamefont {D.~J.}\ \bibnamefont
  {Twitchen}}, \bibinfo {author} {\bibfnamefont {H.}~\bibnamefont {Park}},
  \bibinfo {author} {\bibfnamefont {M.~D.}\ \bibnamefont {Lukin}}, \ and\
  \bibinfo {author} {\bibfnamefont {M.}~\bibnamefont {Loncar}},\ }\href
  {\doibase 10.1021/nl402174g} {\bibfield  {journal} {\bibinfo  {journal} {Nano
  Lett.}\ }\textbf {\bibinfo {volume} {13}},\ \bibinfo {pages} {5791} (\bibinfo
  {year} {2013})}\BibitemShut {NoStop}%
\bibitem [{\citenamefont {Li}\ \emph {et~al.}(2015)\citenamefont {Li},
  \citenamefont {Schr{\"{o}}der}, \citenamefont {Chen}, \citenamefont {Walsh},
  \citenamefont {Bayn}, \citenamefont {Goldstein}, \citenamefont {Gaathon},
  \citenamefont {Trusheim}, \citenamefont {Lu}, \citenamefont {Mower},
  \citenamefont {Cotlet}, \citenamefont {Markham}, \citenamefont {Twitchen},\
  and\ \citenamefont {Englund}}]{Li2014}%
  \BibitemOpen
  \bibfield  {author} {\bibinfo {author} {\bibfnamefont {L.}~\bibnamefont
  {Li}}, \bibinfo {author} {\bibfnamefont {T.}~\bibnamefont {Schr{\"{o}}der}},
  \bibinfo {author} {\bibfnamefont {E.~H.}\ \bibnamefont {Chen}}, \bibinfo
  {author} {\bibfnamefont {M.}~\bibnamefont {Walsh}}, \bibinfo {author}
  {\bibfnamefont {I.}~\bibnamefont {Bayn}}, \bibinfo {author} {\bibfnamefont
  {J.}~\bibnamefont {Goldstein}}, \bibinfo {author} {\bibfnamefont
  {O.}~\bibnamefont {Gaathon}}, \bibinfo {author} {\bibfnamefont {M.~E.}\
  \bibnamefont {Trusheim}}, \bibinfo {author} {\bibfnamefont {M.}~\bibnamefont
  {Lu}}, \bibinfo {author} {\bibfnamefont {J.}~\bibnamefont {Mower}}, \bibinfo
  {author} {\bibfnamefont {M.}~\bibnamefont {Cotlet}}, \bibinfo {author}
  {\bibfnamefont {M.~L.}\ \bibnamefont {Markham}}, \bibinfo {author}
  {\bibfnamefont {D.~J.}\ \bibnamefont {Twitchen}}, \ and\ \bibinfo {author}
  {\bibfnamefont {D.}~\bibnamefont {Englund}},\ }\href {\doibase
  10.1038/ncomms7173} {\bibfield  {journal} {\bibinfo  {journal} {Nat.
  Commun.}\ }\textbf {\bibinfo {volume} {6}},\ \bibinfo {pages} {6173}
  (\bibinfo {year} {2015})}\BibitemShut {NoStop}%
\bibitem [{\citenamefont {Barbour}\ \emph {et~al.}(2011)\citenamefont
  {Barbour}, \citenamefont {Dalgarno}, \citenamefont {Curran}, \citenamefont
  {Nowak}, \citenamefont {Baker}, \citenamefont {Hall}, \citenamefont {Stoltz},
  \citenamefont {Petroff},\ and\ \citenamefont {Warburton}}]{Barbour2011}%
  \BibitemOpen
  \bibfield  {author} {\bibinfo {author} {\bibfnamefont {R.~J.}\ \bibnamefont
  {Barbour}}, \bibinfo {author} {\bibfnamefont {P.~A.}\ \bibnamefont
  {Dalgarno}}, \bibinfo {author} {\bibfnamefont {A.}~\bibnamefont {Curran}},
  \bibinfo {author} {\bibfnamefont {K.~M.}\ \bibnamefont {Nowak}}, \bibinfo
  {author} {\bibfnamefont {H.~J.}\ \bibnamefont {Baker}}, \bibinfo {author}
  {\bibfnamefont {D.~R.}\ \bibnamefont {Hall}}, \bibinfo {author}
  {\bibfnamefont {N.~G.}\ \bibnamefont {Stoltz}}, \bibinfo {author}
  {\bibfnamefont {P.~M.}\ \bibnamefont {Petroff}}, \ and\ \bibinfo {author}
  {\bibfnamefont {R.~J.}\ \bibnamefont {Warburton}},\ }\href {\doibase
  10.1063/1.3632057} {\bibfield  {journal} {\bibinfo  {journal} {J. Appl.
  Phys.}\ }\textbf {\bibinfo {volume} {110}},\ \bibinfo {pages} {053107}
  (\bibinfo {year} {2011})}\BibitemShut {NoStop}%
\bibitem [{\citenamefont {Greuter}\ \emph {et~al.}(2014)\citenamefont
  {Greuter}, \citenamefont {Starosielec}, \citenamefont {Najer}, \citenamefont
  {Ludwig}, \citenamefont {Duempelmann}, \citenamefont {Rohner},\ and\
  \citenamefont {Warburton}}]{Greuter2014}%
  \BibitemOpen
  \bibfield  {author} {\bibinfo {author} {\bibfnamefont {L.}~\bibnamefont
  {Greuter}}, \bibinfo {author} {\bibfnamefont {S.}~\bibnamefont
  {Starosielec}}, \bibinfo {author} {\bibfnamefont {D.}~\bibnamefont {Najer}},
  \bibinfo {author} {\bibfnamefont {A.}~\bibnamefont {Ludwig}}, \bibinfo
  {author} {\bibfnamefont {L.}~\bibnamefont {Duempelmann}}, \bibinfo {author}
  {\bibfnamefont {D.}~\bibnamefont {Rohner}}, \ and\ \bibinfo {author}
  {\bibfnamefont {R.~J.}\ \bibnamefont {Warburton}},\ }\href {\doibase
  10.1063/1.4896415} {\bibfield  {journal} {\bibinfo  {journal} {Appl. Phys.
  Lett.}\ }\textbf {\bibinfo {volume} {105}},\ \bibinfo {pages} {121105}
  (\bibinfo {year} {2014})}\BibitemShut {NoStop}%
\bibitem [{\citenamefont {Greuter}\ \emph {et~al.}(2015)\citenamefont
  {Greuter}, \citenamefont {Starosielec}, \citenamefont {Kuhlmann},\ and\
  \citenamefont {Warburton}}]{Greuter2015}%
  \BibitemOpen
  \bibfield  {author} {\bibinfo {author} {\bibfnamefont {L.}~\bibnamefont
  {Greuter}}, \bibinfo {author} {\bibfnamefont {S.}~\bibnamefont
  {Starosielec}}, \bibinfo {author} {\bibfnamefont {A.~V.}\ \bibnamefont
  {Kuhlmann}}, \ and\ \bibinfo {author} {\bibfnamefont {R.~J.}\ \bibnamefont
  {Warburton}},\ }\href {\doibase 10.1103/PhysRevB.92.045302} {\bibfield
  {journal} {\bibinfo  {journal} {Phys. Rev. B}\ }\textbf {\bibinfo {volume}
  {92}},\ \bibinfo {pages} {045302} (\bibinfo {year} {2015})}\BibitemShut
  {NoStop}%
\bibitem [{\citenamefont {Albrecht}\ \emph {et~al.}(2013)\citenamefont
  {Albrecht}, \citenamefont {Bommer}, \citenamefont {Deutsch}, \citenamefont
  {Reichel},\ and\ \citenamefont {Becher}}]{Albrecht2013}%
  \BibitemOpen
  \bibfield  {author} {\bibinfo {author} {\bibfnamefont {R.}~\bibnamefont
  {Albrecht}}, \bibinfo {author} {\bibfnamefont {A.}~\bibnamefont {Bommer}},
  \bibinfo {author} {\bibfnamefont {C.}~\bibnamefont {Deutsch}}, \bibinfo
  {author} {\bibfnamefont {J.}~\bibnamefont {Reichel}}, \ and\ \bibinfo
  {author} {\bibfnamefont {C.}~\bibnamefont {Becher}},\ }\href {\doibase
  10.1103/PhysRevLett.110.243602} {\bibfield  {journal} {\bibinfo  {journal}
  {Phys. Rev. Lett.}\ }\textbf {\bibinfo {volume} {110}},\ \bibinfo {pages}
  {243602} (\bibinfo {year} {2013})}\BibitemShut {NoStop}%
\bibitem [{\citenamefont {Johnson}\ \emph {et~al.}(2015)\citenamefont
  {Johnson}, \citenamefont {Dolan}, \citenamefont {Grange}, \citenamefont
  {Trichet}, \citenamefont {Hornecker}, \citenamefont {Chen}, \citenamefont
  {Weng}, \citenamefont {Hughes}, \citenamefont {Auff{\`{e}}ves},\ and\
  \citenamefont {Smith}}]{Johnson2015}%
  \BibitemOpen
  \bibfield  {author} {\bibinfo {author} {\bibfnamefont {S.}~\bibnamefont
  {Johnson}}, \bibinfo {author} {\bibfnamefont {P.~R.}\ \bibnamefont {Dolan}},
  \bibinfo {author} {\bibfnamefont {T.}~\bibnamefont {Grange}}, \bibinfo
  {author} {\bibfnamefont {A.~A.~P.}\ \bibnamefont {Trichet}}, \bibinfo
  {author} {\bibfnamefont {G.}~\bibnamefont {Hornecker}}, \bibinfo {author}
  {\bibfnamefont {Y.~C.}\ \bibnamefont {Chen}}, \bibinfo {author}
  {\bibfnamefont {L.}~\bibnamefont {Weng}}, \bibinfo {author} {\bibfnamefont
  {G.~M.}\ \bibnamefont {Hughes}}, \bibinfo {author} {\bibfnamefont
  {A.}~\bibnamefont {Auff{\`{e}}ves}}, \ and\ \bibinfo {author} {\bibfnamefont
  {J.~M.}\ \bibnamefont {Smith}},\ }\href {\doibase
  10.1088/1367-2630/17/12/122003} {\bibfield  {journal} {\bibinfo  {journal}
  {New J. Phys.}\ }\textbf {\bibinfo {volume} {17}},\ \bibinfo {pages} {122003}
  (\bibinfo {year} {2015})}\BibitemShut {NoStop}%
\bibitem [{\citenamefont {Kaupp}\ \emph {et~al.}(2016)\citenamefont {Kaupp},
  \citenamefont {H{\"{u}}mmer}, \citenamefont {Mader}, \citenamefont
  {Schlederer}, \citenamefont {Benedikter}, \citenamefont {Haeusser},
  \citenamefont {Chang}, \citenamefont {Fedder}, \citenamefont {H{\"{a}}nsch},\
  and\ \citenamefont {Hunger}}]{Kaupp2016}%
  \BibitemOpen
  \bibfield  {author} {\bibinfo {author} {\bibfnamefont {H.}~\bibnamefont
  {Kaupp}}, \bibinfo {author} {\bibfnamefont {T.}~\bibnamefont {H{\"{u}}mmer}},
  \bibinfo {author} {\bibfnamefont {M.}~\bibnamefont {Mader}}, \bibinfo
  {author} {\bibfnamefont {B.}~\bibnamefont {Schlederer}}, \bibinfo {author}
  {\bibfnamefont {J.}~\bibnamefont {Benedikter}}, \bibinfo {author}
  {\bibfnamefont {P.}~\bibnamefont {Haeusser}}, \bibinfo {author}
  {\bibfnamefont {H.-C.}\ \bibnamefont {Chang}}, \bibinfo {author}
  {\bibfnamefont {H.}~\bibnamefont {Fedder}}, \bibinfo {author} {\bibfnamefont
  {T.~W.}\ \bibnamefont {H{\"{a}}nsch}}, \ and\ \bibinfo {author}
  {\bibfnamefont {D.}~\bibnamefont {Hunger}},\ }\href {\doibase
  10.1103/PhysRevApplied.6.054010} {\bibfield  {journal} {\bibinfo  {journal}
  {Phys. Rev. Appl.}\ }\textbf {\bibinfo {volume} {6}},\ \bibinfo {pages}
  {054010} (\bibinfo {year} {2016})}\BibitemShut {NoStop}%
\bibitem [{\citenamefont {Benedikter}\ \emph {et~al.}(2015)\citenamefont
  {Benedikter}, \citenamefont {H{\"{u}}mmer}, \citenamefont {Mader},
  \citenamefont {Schlederer}, \citenamefont {Reichel}, \citenamefont
  {H{\"{a}}nsch},\ and\ \citenamefont {Hunger}}]{Benedikter2017}%
  \BibitemOpen
  \bibfield  {author} {\bibinfo {author} {\bibfnamefont {J.}~\bibnamefont
  {Benedikter}}, \bibinfo {author} {\bibfnamefont {T.}~\bibnamefont
  {H{\"{u}}mmer}}, \bibinfo {author} {\bibfnamefont {M.}~\bibnamefont {Mader}},
  \bibinfo {author} {\bibfnamefont {B.}~\bibnamefont {Schlederer}}, \bibinfo
  {author} {\bibfnamefont {J.}~\bibnamefont {Reichel}}, \bibinfo {author}
  {\bibfnamefont {T.~W.}\ \bibnamefont {H{\"{a}}nsch}}, \ and\ \bibinfo
  {author} {\bibfnamefont {D.}~\bibnamefont {Hunger}},\ }\href
  {http://arxiv.org/abs/1502.01532} {\bibfield  {journal} {\bibinfo  {journal}
  {arXiv: 1502.01532}\ } (\bibinfo {year} {2015})}\BibitemShut {NoStop}%
\bibitem [{\citenamefont {Greentree}\ \emph {et~al.}(2006)\citenamefont
  {Greentree}, \citenamefont {Olivero}, \citenamefont {Draganski},
  \citenamefont {Trajkov}, \citenamefont {Rabeau}, \citenamefont {Reichart},
  \citenamefont {Gibson}, \citenamefont {Rubanov}, \citenamefont {Huntington},
  \citenamefont {Jamieson},\ and\ \citenamefont {Prawer}}]{Greentree2006}%
  \BibitemOpen
  \bibfield  {author} {\bibinfo {author} {\bibfnamefont {A.~D.}\ \bibnamefont
  {Greentree}}, \bibinfo {author} {\bibfnamefont {P.}~\bibnamefont {Olivero}},
  \bibinfo {author} {\bibfnamefont {M.}~\bibnamefont {Draganski}}, \bibinfo
  {author} {\bibfnamefont {E.}~\bibnamefont {Trajkov}}, \bibinfo {author}
  {\bibfnamefont {J.~R.}\ \bibnamefont {Rabeau}}, \bibinfo {author}
  {\bibfnamefont {P.}~\bibnamefont {Reichart}}, \bibinfo {author}
  {\bibfnamefont {B.~C.}\ \bibnamefont {Gibson}}, \bibinfo {author}
  {\bibfnamefont {S.}~\bibnamefont {Rubanov}}, \bibinfo {author} {\bibfnamefont
  {S.~T.}\ \bibnamefont {Huntington}}, \bibinfo {author} {\bibfnamefont
  {D.~N.}\ \bibnamefont {Jamieson}}, \ and\ \bibinfo {author} {\bibfnamefont
  {S.}~\bibnamefont {Prawer}},\ }\href {\doibase 10.1088/0953-8984/18/21/S09}
  {\bibfield  {journal} {\bibinfo  {journal} {J. Phys. Condens. Matter}\
  }\textbf {\bibinfo {volume} {18}},\ \bibinfo {pages} {S825} (\bibinfo {year}
  {2006})}\BibitemShut {NoStop}%
\bibitem [{\citenamefont {Schr{\"{o}}der}\ \emph {et~al.}(2016)\citenamefont
  {Schr{\"{o}}der}, \citenamefont {Mouradian}, \citenamefont {Zheng},
  \citenamefont {Trusheim}, \citenamefont {Walsh}, \citenamefont {Chen},
  \citenamefont {Li}, \citenamefont {Bayn},\ and\ \citenamefont
  {Englund}}]{Schroder2016}%
  \BibitemOpen
  \bibfield  {author} {\bibinfo {author} {\bibfnamefont {T.}~\bibnamefont
  {Schr{\"{o}}der}}, \bibinfo {author} {\bibfnamefont {S.~L.}\ \bibnamefont
  {Mouradian}}, \bibinfo {author} {\bibfnamefont {J.}~\bibnamefont {Zheng}},
  \bibinfo {author} {\bibfnamefont {M.~E.}\ \bibnamefont {Trusheim}}, \bibinfo
  {author} {\bibfnamefont {M.}~\bibnamefont {Walsh}}, \bibinfo {author}
  {\bibfnamefont {E.~H.}\ \bibnamefont {Chen}}, \bibinfo {author}
  {\bibfnamefont {L.}~\bibnamefont {Li}}, \bibinfo {author} {\bibfnamefont
  {I.}~\bibnamefont {Bayn}}, \ and\ \bibinfo {author} {\bibfnamefont
  {D.}~\bibnamefont {Englund}},\ }\href {\doibase 10.1364/JOSAB.33.000B65}
  {\bibfield  {journal} {\bibinfo  {journal} {J. Opt. Soc. Am. B}\ }\textbf
  {\bibinfo {volume} {33}},\ \bibinfo {pages} {B65} (\bibinfo {year}
  {2016})}\BibitemShut {NoStop}%
\bibitem [{\citenamefont {Chu}\ \emph {et~al.}(2014)\citenamefont {Chu},
  \citenamefont {de~Leon}, \citenamefont {Shields}, \citenamefont {Hausmann},
  \citenamefont {Evans}, \citenamefont {Togan}, \citenamefont {Burek},
  \citenamefont {Markham}, \citenamefont {Stacey}, \citenamefont {Zibrov},
  \citenamefont {Yacoby}, \citenamefont {Twitchen}, \citenamefont {Loncar},
  \citenamefont {Park}, \citenamefont {Maletinsky},\ and\ \citenamefont
  {Lukin}}]{Chu2014b}%
  \BibitemOpen
  \bibfield  {author} {\bibinfo {author} {\bibfnamefont {Y.}~\bibnamefont
  {Chu}}, \bibinfo {author} {\bibfnamefont {N.~P.}\ \bibnamefont {de~Leon}},
  \bibinfo {author} {\bibfnamefont {B.~J.}\ \bibnamefont {Shields}}, \bibinfo
  {author} {\bibfnamefont {B.}~\bibnamefont {Hausmann}}, \bibinfo {author}
  {\bibfnamefont {R.}~\bibnamefont {Evans}}, \bibinfo {author} {\bibfnamefont
  {E.}~\bibnamefont {Togan}}, \bibinfo {author} {\bibfnamefont {M.~J.}\
  \bibnamefont {Burek}}, \bibinfo {author} {\bibfnamefont {M.}~\bibnamefont
  {Markham}}, \bibinfo {author} {\bibfnamefont {A.}~\bibnamefont {Stacey}},
  \bibinfo {author} {\bibfnamefont {A.~S.}\ \bibnamefont {Zibrov}}, \bibinfo
  {author} {\bibfnamefont {A.}~\bibnamefont {Yacoby}}, \bibinfo {author}
  {\bibfnamefont {D.~J.}\ \bibnamefont {Twitchen}}, \bibinfo {author}
  {\bibfnamefont {M.}~\bibnamefont {Loncar}}, \bibinfo {author} {\bibfnamefont
  {H.}~\bibnamefont {Park}}, \bibinfo {author} {\bibfnamefont {P.}~\bibnamefont
  {Maletinsky}}, \ and\ \bibinfo {author} {\bibfnamefont {M.~D.}\ \bibnamefont
  {Lukin}},\ }\href {\doibase 10.1021/nl404836p} {\bibfield  {journal}
  {\bibinfo  {journal} {Nano Lett.}\ }\textbf {\bibinfo {volume} {14}},\
  \bibinfo {pages} {1982} (\bibinfo {year} {2014})}\BibitemShut {NoStop}%
\bibitem [{\citenamefont {Appel}\ \emph {et~al.}(2016)\citenamefont {Appel},
  \citenamefont {Neu}, \citenamefont {Ganzhorn}, \citenamefont {Barfuss},
  \citenamefont {Batzer}, \citenamefont {Gratz}, \citenamefont
  {Tsch{\"{o}}pe},\ and\ \citenamefont {Maletinsky}}]{Appel2016}%
  \BibitemOpen
  \bibfield  {author} {\bibinfo {author} {\bibfnamefont {P.}~\bibnamefont
  {Appel}}, \bibinfo {author} {\bibfnamefont {E.}~\bibnamefont {Neu}}, \bibinfo
  {author} {\bibfnamefont {M.}~\bibnamefont {Ganzhorn}}, \bibinfo {author}
  {\bibfnamefont {A.}~\bibnamefont {Barfuss}}, \bibinfo {author} {\bibfnamefont
  {M.}~\bibnamefont {Batzer}}, \bibinfo {author} {\bibfnamefont
  {M.}~\bibnamefont {Gratz}}, \bibinfo {author} {\bibfnamefont
  {A.}~\bibnamefont {Tsch{\"{o}}pe}}, \ and\ \bibinfo {author} {\bibfnamefont
  {P.}~\bibnamefont {Maletinsky}},\ }\href {\doibase 10.1063/1.4952953}
  {\bibfield  {journal} {\bibinfo  {journal} {Rev. Sci. Instrum.}\ }\textbf
  {\bibinfo {volume} {87}},\ \bibinfo {pages} {063703} (\bibinfo {year}
  {2016})}\BibitemShut {NoStop}%
\bibitem [{\citenamefont {Maletinsky}\ \emph {et~al.}(2012)\citenamefont
  {Maletinsky}, \citenamefont {Hong}, \citenamefont {Grinolds}, \citenamefont
  {Hausmann}, \citenamefont {Lukin}, \citenamefont {Walsworth}, \citenamefont
  {Loncar},\ and\ \citenamefont {Yacoby}}]{Maletinsky2012}%
  \BibitemOpen
  \bibfield  {author} {\bibinfo {author} {\bibfnamefont {P.}~\bibnamefont
  {Maletinsky}}, \bibinfo {author} {\bibfnamefont {S.}~\bibnamefont {Hong}},
  \bibinfo {author} {\bibfnamefont {M.~S.}\ \bibnamefont {Grinolds}}, \bibinfo
  {author} {\bibfnamefont {B.}~\bibnamefont {Hausmann}}, \bibinfo {author}
  {\bibfnamefont {M.~D.}\ \bibnamefont {Lukin}}, \bibinfo {author}
  {\bibfnamefont {R.~L.}\ \bibnamefont {Walsworth}}, \bibinfo {author}
  {\bibfnamefont {M.}~\bibnamefont {Loncar}}, \ and\ \bibinfo {author}
  {\bibfnamefont {A.}~\bibnamefont {Yacoby}},\ }\href {\doibase
  10.1038/nnano.2012.50} {\bibfield  {journal} {\bibinfo  {journal} {Nat.
  Nanotechnol.}\ }\textbf {\bibinfo {volume} {7}},\ \bibinfo {pages} {320}
  (\bibinfo {year} {2012})}\BibitemShut {NoStop}%
\bibitem [{\citenamefont {Riedel}\ \emph {et~al.}(2014)\citenamefont {Riedel},
  \citenamefont {Rohner}, \citenamefont {Ganzhorn}, \citenamefont {Kaldewey},
  \citenamefont {Appel}, \citenamefont {Neu}, \citenamefont {Warburton},\ and\
  \citenamefont {Maletinsky}}]{Riedel2014}%
  \BibitemOpen
  \bibfield  {author} {\bibinfo {author} {\bibfnamefont {D.}~\bibnamefont
  {Riedel}}, \bibinfo {author} {\bibfnamefont {D.}~\bibnamefont {Rohner}},
  \bibinfo {author} {\bibfnamefont {M.}~\bibnamefont {Ganzhorn}}, \bibinfo
  {author} {\bibfnamefont {T.}~\bibnamefont {Kaldewey}}, \bibinfo {author}
  {\bibfnamefont {P.}~\bibnamefont {Appel}}, \bibinfo {author} {\bibfnamefont
  {E.}~\bibnamefont {Neu}}, \bibinfo {author} {\bibfnamefont {R.~J.}\
  \bibnamefont {Warburton}}, \ and\ \bibinfo {author} {\bibfnamefont
  {P.}~\bibnamefont {Maletinsky}},\ }\href {\doibase
  10.1103/PhysRevApplied.2.064011} {\bibfield  {journal} {\bibinfo  {journal}
  {Phys. Rev. Appl.}\ }\textbf {\bibinfo {volume} {2}},\ \bibinfo {pages}
  {064011} (\bibinfo {year} {2014})}\BibitemShut {NoStop}%
\bibitem [{\citenamefont {Hunger}\ \emph {et~al.}(2012)\citenamefont {Hunger},
  \citenamefont {Deutsch}, \citenamefont {Barbour}, \citenamefont {Warburton},\
  and\ \citenamefont {Reichel}}]{Hunger2012}%
  \BibitemOpen
  \bibfield  {author} {\bibinfo {author} {\bibfnamefont {D.}~\bibnamefont
  {Hunger}}, \bibinfo {author} {\bibfnamefont {C.}~\bibnamefont {Deutsch}},
  \bibinfo {author} {\bibfnamefont {R.~J.}\ \bibnamefont {Barbour}}, \bibinfo
  {author} {\bibfnamefont {R.~J.}\ \bibnamefont {Warburton}}, \ and\ \bibinfo
  {author} {\bibfnamefont {J.}~\bibnamefont {Reichel}},\ }\href {\doibase
  10.1063/1.3679721} {\bibfield  {journal} {\bibinfo  {journal} {AIP Adv.}\
  }\textbf {\bibinfo {volume} {2}},\ \bibinfo {pages} {012119} (\bibinfo {year}
  {2012})}\BibitemShut {NoStop}%
\bibitem [{\citenamefont {Faraon}\ \emph {et~al.}(2011)\citenamefont {Faraon},
  \citenamefont {Barclay}, \citenamefont {Santori}, \citenamefont {Fu},\ and\
  \citenamefont {Beausoleil}}]{Faraon2011}%
  \BibitemOpen
  \bibfield  {author} {\bibinfo {author} {\bibfnamefont {A.}~\bibnamefont
  {Faraon}}, \bibinfo {author} {\bibfnamefont {P.~E.}\ \bibnamefont {Barclay}},
  \bibinfo {author} {\bibfnamefont {C.}~\bibnamefont {Santori}}, \bibinfo
  {author} {\bibfnamefont {K.-M.~C.}\ \bibnamefont {Fu}}, \ and\ \bibinfo
  {author} {\bibfnamefont {R.~G.}\ \bibnamefont {Beausoleil}},\ }\href
  {\doibase 10.1038/nphoton.2011.52} {\bibfield  {journal} {\bibinfo  {journal}
  {Nat. Photonics}\ }\textbf {\bibinfo {volume} {5}},\ \bibinfo {pages} {301}
  (\bibinfo {year} {2011})}\BibitemShut {NoStop}%
\bibitem [{\citenamefont {Janitz}\ \emph {et~al.}(2015)\citenamefont {Janitz},
  \citenamefont {Ruf}, \citenamefont {Dimock}, \citenamefont {Bourassa},
  \citenamefont {Sankey},\ and\ \citenamefont {Childress}}]{Janitz2015}%
  \BibitemOpen
  \bibfield  {author} {\bibinfo {author} {\bibfnamefont {E.}~\bibnamefont
  {Janitz}}, \bibinfo {author} {\bibfnamefont {M.}~\bibnamefont {Ruf}},
  \bibinfo {author} {\bibfnamefont {M.}~\bibnamefont {Dimock}}, \bibinfo
  {author} {\bibfnamefont {A.}~\bibnamefont {Bourassa}}, \bibinfo {author}
  {\bibfnamefont {J.}~\bibnamefont {Sankey}}, \ and\ \bibinfo {author}
  {\bibfnamefont {L.}~\bibnamefont {Childress}},\ }\href {\doibase
  10.1103/PhysRevA.92.043844} {\bibfield  {journal} {\bibinfo  {journal} {Phys.
  Rev. A}\ }\textbf {\bibinfo {volume} {92}},\ \bibinfo {pages} {043844}
  (\bibinfo {year} {2015})}\BibitemShut {NoStop}%
\bibitem [{\citenamefont {Radko}\ \emph {et~al.}(2016)\citenamefont {Radko},
  \citenamefont {Boll}, \citenamefont {Israelsen}, \citenamefont {Raatz},
  \citenamefont {Meijer}, \citenamefont {Jelezko}, \citenamefont {Andersen},\
  and\ \citenamefont {Huck}}]{Radko2016}%
  \BibitemOpen
  \bibfield  {author} {\bibinfo {author} {\bibfnamefont {I.~P.}\ \bibnamefont
  {Radko}}, \bibinfo {author} {\bibfnamefont {M.}~\bibnamefont {Boll}},
  \bibinfo {author} {\bibfnamefont {N.~M.}\ \bibnamefont {Israelsen}}, \bibinfo
  {author} {\bibfnamefont {N.}~\bibnamefont {Raatz}}, \bibinfo {author}
  {\bibfnamefont {J.}~\bibnamefont {Meijer}}, \bibinfo {author} {\bibfnamefont
  {F.}~\bibnamefont {Jelezko}}, \bibinfo {author} {\bibfnamefont {U.~L.}\
  \bibnamefont {Andersen}}, \ and\ \bibinfo {author} {\bibfnamefont
  {A.}~\bibnamefont {Huck}},\ }\href {\doibase 10.1364/OE.24.027715} {\bibfield
   {journal} {\bibinfo  {journal} {Opt. Express}\ }\textbf {\bibinfo {volume}
  {24}},\ \bibinfo {pages} {27715} (\bibinfo {year} {2016})}\BibitemShut
  {NoStop}%
\bibitem [{\citenamefont {Najer}\ \emph {et~al.}(2017)\citenamefont {Najer},
  \citenamefont {Renggli}, \citenamefont {Riedel}, \citenamefont
  {Starosielec},\ and\ \citenamefont {Warburton}}]{Najer2017}%
  \BibitemOpen
  \bibfield  {author} {\bibinfo {author} {\bibfnamefont {D.}~\bibnamefont
  {Najer}}, \bibinfo {author} {\bibfnamefont {M.}~\bibnamefont {Renggli}},
  \bibinfo {author} {\bibfnamefont {D.}~\bibnamefont {Riedel}}, \bibinfo
  {author} {\bibfnamefont {S.}~\bibnamefont {Starosielec}}, \ and\ \bibinfo
  {author} {\bibfnamefont {R.~J.}\ \bibnamefont {Warburton}},\ }\href {\doibase
  10.1063/1.4973458} {\bibfield  {journal} {\bibinfo  {journal} {Appl. Phys.
  Lett.}\ }\textbf {\bibinfo {volume} {110}},\ \bibinfo {pages} {011101}
  (\bibinfo {year} {2017})}\BibitemShut {NoStop}%
\bibitem [{\citenamefont {Bogdanovic}\ \emph {et~al.}(2016)\citenamefont
  {Bogdanovic}, \citenamefont {van Dam}, \citenamefont {Bonato}, \citenamefont
  {Coenen}, \citenamefont {Zwerver}, \citenamefont {Hensen}, \citenamefont
  {Liddy}, \citenamefont {Fink}, \citenamefont {Reiserer}, \citenamefont
  {Loncar},\ and\ \citenamefont {Hanson}}]{Bogdanovic2017}%
  \BibitemOpen
  \bibfield  {author} {\bibinfo {author} {\bibfnamefont {S.}~\bibnamefont
  {Bogdanovic}}, \bibinfo {author} {\bibfnamefont {S.~B.}\ \bibnamefont {van
  Dam}}, \bibinfo {author} {\bibfnamefont {C.}~\bibnamefont {Bonato}}, \bibinfo
  {author} {\bibfnamefont {L.~C.}\ \bibnamefont {Coenen}}, \bibinfo {author}
  {\bibfnamefont {A.~J.}\ \bibnamefont {Zwerver}}, \bibinfo {author}
  {\bibfnamefont {B.}~\bibnamefont {Hensen}}, \bibinfo {author} {\bibfnamefont
  {M.~S.~Z.}\ \bibnamefont {Liddy}}, \bibinfo {author} {\bibfnamefont
  {T.}~\bibnamefont {Fink}}, \bibinfo {author} {\bibfnamefont {A.}~\bibnamefont
  {Reiserer}}, \bibinfo {author} {\bibfnamefont {M.}~\bibnamefont {Loncar}}, \
  and\ \bibinfo {author} {\bibfnamefont {R.}~\bibnamefont {Hanson}},\ }\href
  {http://arxiv.org/abs/1612.02164} {\bibfield  {journal} {\bibinfo  {journal}
  {arXiv: 1612.02164}\ } (\bibinfo {year} {2016})}\BibitemShut {NoStop}%
\bibitem [{\citenamefont {Aharonovich}\ and\ \citenamefont
  {Neu}(2014)}]{Aharonovich2014}%
  \BibitemOpen
  \bibfield  {author} {\bibinfo {author} {\bibfnamefont {I.}~\bibnamefont
  {Aharonovich}}\ and\ \bibinfo {author} {\bibfnamefont {E.}~\bibnamefont
  {Neu}},\ }\href {\doibase 10.1002/adom.201400189} {\bibfield  {journal}
  {\bibinfo  {journal} {Adv. Opt. Mater.}\ }\textbf {\bibinfo {volume} {2}},\
  \bibinfo {pages} {911} (\bibinfo {year} {2014})}\BibitemShut {NoStop}%
\bibitem [{\citenamefont {Siyushev}\ \emph {et~al.}(2016)\citenamefont
  {Siyushev}, \citenamefont {Metsch}, \citenamefont {Ijaz}, \citenamefont
  {Binder}, \citenamefont {Bhaskar}, \citenamefont {Sukachev}, \citenamefont
  {Sipahigil}, \citenamefont {Evans}, \citenamefont {Nguyen}, \citenamefont
  {Lukin}, \citenamefont {Hemmer}, \citenamefont {Palyanov}, \citenamefont
  {Kupriyanov}, \citenamefont {Borzdov}, \citenamefont {Rogers},\ and\
  \citenamefont {Jelezko}}]{Siyushev2016}%
  \BibitemOpen
  \bibfield  {author} {\bibinfo {author} {\bibfnamefont {P.}~\bibnamefont
  {Siyushev}}, \bibinfo {author} {\bibfnamefont {M.~H.}\ \bibnamefont
  {Metsch}}, \bibinfo {author} {\bibfnamefont {A.}~\bibnamefont {Ijaz}},
  \bibinfo {author} {\bibfnamefont {J.~M.}\ \bibnamefont {Binder}}, \bibinfo
  {author} {\bibfnamefont {M.~K.}\ \bibnamefont {Bhaskar}}, \bibinfo {author}
  {\bibfnamefont {D.~D.}\ \bibnamefont {Sukachev}}, \bibinfo {author}
  {\bibfnamefont {A.}~\bibnamefont {Sipahigil}}, \bibinfo {author}
  {\bibfnamefont {R.~E.}\ \bibnamefont {Evans}}, \bibinfo {author}
  {\bibfnamefont {C.~T.}\ \bibnamefont {Nguyen}}, \bibinfo {author}
  {\bibfnamefont {M.~D.}\ \bibnamefont {Lukin}}, \bibinfo {author}
  {\bibfnamefont {P.~R.}\ \bibnamefont {Hemmer}}, \bibinfo {author}
  {\bibfnamefont {Y.~N.}\ \bibnamefont {Palyanov}}, \bibinfo {author}
  {\bibfnamefont {I.~N.}\ \bibnamefont {Kupriyanov}}, \bibinfo {author}
  {\bibfnamefont {Y.~M.}\ \bibnamefont {Borzdov}}, \bibinfo {author}
  {\bibfnamefont {L.~J.}\ \bibnamefont {Rogers}}, \ and\ \bibinfo {author}
  {\bibfnamefont {F.}~\bibnamefont {Jelezko}},\ }\href
  {http://arxiv.org/abs/1612.02947} {\bibfield  {journal} {\bibinfo  {journal}
  {arXiV: 1612.02947}\ } (\bibinfo {year} {2016})}\BibitemShut {NoStop}%
\bibitem [{\citenamefont {Bhaskar}\ \emph {et~al.}(2016)\citenamefont
  {Bhaskar}, \citenamefont {Sukachev}, \citenamefont {Sipahigil}, \citenamefont
  {Evans}, \citenamefont {Burek}, \citenamefont {Nguyen}, \citenamefont
  {Rogers}, \citenamefont {Siyushev}, \citenamefont {Metsch}, \citenamefont
  {Park}, \citenamefont {Jelezko}, \citenamefont {Lon{\v{c}}ar},\ and\
  \citenamefont {Lukin}}]{Bhaskar2016}%
  \BibitemOpen
  \bibfield  {author} {\bibinfo {author} {\bibfnamefont {M.~K.}\ \bibnamefont
  {Bhaskar}}, \bibinfo {author} {\bibfnamefont {D.~D.}\ \bibnamefont
  {Sukachev}}, \bibinfo {author} {\bibfnamefont {A.}~\bibnamefont {Sipahigil}},
  \bibinfo {author} {\bibfnamefont {R.~E.}\ \bibnamefont {Evans}}, \bibinfo
  {author} {\bibfnamefont {M.~J.}\ \bibnamefont {Burek}}, \bibinfo {author}
  {\bibfnamefont {C.~T.}\ \bibnamefont {Nguyen}}, \bibinfo {author}
  {\bibfnamefont {L.~J.}\ \bibnamefont {Rogers}}, \bibinfo {author}
  {\bibfnamefont {P.}~\bibnamefont {Siyushev}}, \bibinfo {author}
  {\bibfnamefont {M.~H.}\ \bibnamefont {Metsch}}, \bibinfo {author}
  {\bibfnamefont {H.}~\bibnamefont {Park}}, \bibinfo {author} {\bibfnamefont
  {F.}~\bibnamefont {Jelezko}}, \bibinfo {author} {\bibfnamefont
  {M.}~\bibnamefont {Lon{\v{c}}ar}}, \ and\ \bibinfo {author} {\bibfnamefont
  {M.~D.}\ \bibnamefont {Lukin}},\ }\href {http://arxiv.org/abs/1612.03036}
  {\bibfield  {journal} {\bibinfo  {journal} {arXiV: 1612.03036}\ } (\bibinfo
  {year} {2016})}\BibitemShut {NoStop}%
\bibitem [{\citenamefont {Koehl}\ \emph {et~al.}(2011)\citenamefont {Koehl},
  \citenamefont {Buckley}, \citenamefont {Heremans}, \citenamefont {Calusine},\
  and\ \citenamefont {Awschalom}}]{Koehl2011}%
  \BibitemOpen
  \bibfield  {author} {\bibinfo {author} {\bibfnamefont {W.~F.}\ \bibnamefont
  {Koehl}}, \bibinfo {author} {\bibfnamefont {B.~B.}\ \bibnamefont {Buckley}},
  \bibinfo {author} {\bibfnamefont {F.~J.}\ \bibnamefont {Heremans}}, \bibinfo
  {author} {\bibfnamefont {G.}~\bibnamefont {Calusine}}, \ and\ \bibinfo
  {author} {\bibfnamefont {D.~D.}\ \bibnamefont {Awschalom}},\ }\href {\doibase
  10.1038/nature10562} {\bibfield  {journal} {\bibinfo  {journal} {Nature}\
  }\textbf {\bibinfo {volume} {479}},\ \bibinfo {pages} {84} (\bibinfo {year}
  {2011})}\BibitemShut {NoStop}%
\bibitem [{\citenamefont {Castelletto}\ \emph {et~al.}(2014)\citenamefont
  {Castelletto}, \citenamefont {Johnson}, \citenamefont {Iv{\'{a}}dy},
  \citenamefont {Stavrias}, \citenamefont {Umeda}, \citenamefont {Gali},\ and\
  \citenamefont {Ohshima}}]{Castelletto2014}%
  \BibitemOpen
  \bibfield  {author} {\bibinfo {author} {\bibfnamefont {S.}~\bibnamefont
  {Castelletto}}, \bibinfo {author} {\bibfnamefont {B.~C.}\ \bibnamefont
  {Johnson}}, \bibinfo {author} {\bibfnamefont {V.}~\bibnamefont
  {Iv{\'{a}}dy}}, \bibinfo {author} {\bibfnamefont {N.}~\bibnamefont
  {Stavrias}}, \bibinfo {author} {\bibfnamefont {T.}~\bibnamefont {Umeda}},
  \bibinfo {author} {\bibfnamefont {A.}~\bibnamefont {Gali}}, \ and\ \bibinfo
  {author} {\bibfnamefont {T.}~\bibnamefont {Ohshima}},\ }\href {\doibase
  10.1038/nmat3806} {\bibfield  {journal} {\bibinfo  {journal} {Nat. Mater.}\
  }\textbf {\bibinfo {volume} {13}},\ \bibinfo {pages} {151} (\bibinfo {year}
  {2014})}\BibitemShut {NoStop}%
\bibitem [{\citenamefont {Riedel}\ \emph {et~al.}(2012)\citenamefont {Riedel},
  \citenamefont {Fuchs}, \citenamefont {Kraus}, \citenamefont {V{\"{a}}th},
  \citenamefont {Sperlich}, \citenamefont {Dyakonov}, \citenamefont
  {Soltamova}, \citenamefont {Baranov}, \citenamefont {Ilyin},\ and\
  \citenamefont {Astakhov}}]{Riedel2012}%
  \BibitemOpen
  \bibfield  {author} {\bibinfo {author} {\bibfnamefont {D.}~\bibnamefont
  {Riedel}}, \bibinfo {author} {\bibfnamefont {F.}~\bibnamefont {Fuchs}},
  \bibinfo {author} {\bibfnamefont {H.}~\bibnamefont {Kraus}}, \bibinfo
  {author} {\bibfnamefont {S.}~\bibnamefont {V{\"{a}}th}}, \bibinfo {author}
  {\bibfnamefont {A.}~\bibnamefont {Sperlich}}, \bibinfo {author}
  {\bibfnamefont {V.}~\bibnamefont {Dyakonov}}, \bibinfo {author}
  {\bibfnamefont {A.~A.}\ \bibnamefont {Soltamova}}, \bibinfo {author}
  {\bibfnamefont {P.~G.}\ \bibnamefont {Baranov}}, \bibinfo {author}
  {\bibfnamefont {V.~A.}\ \bibnamefont {Ilyin}}, \ and\ \bibinfo {author}
  {\bibfnamefont {G.~V.}\ \bibnamefont {Astakhov}},\ }\href {\doibase
  10.1103/PhysRevLett.109.226402} {\bibfield  {journal} {\bibinfo  {journal}
  {Phys. Rev. Lett.}\ }\textbf {\bibinfo {volume} {109}},\ \bibinfo {pages}
  {226402} (\bibinfo {year} {2012})}\BibitemShut {NoStop}%
\end{thebibliography}

%

\end{document}